\documentclass[prd,twocolumn,showpacs,floatfix,nofootinbib]{revtex4}

\usepackage{amsfonts}
\usepackage[colorlinks=true,citecolor=blue,linkcolor=blue,urlcolor=blue]{hyperref}
\usepackage{color,xcolor}
\usepackage{graphicx,amsfonts,multirow}
\usepackage{amsmath}
\usepackage{amsmath,amssymb,graphicx,latexsym,subfigure}
\usepackage{threeparttable,multirow,makecell,tabularx}
\usepackage{ulem}

\newcommand{\DS}[1]{/\!\!\!#1}

\allowdisplaybreaks

\begin{document}

\title{Light-cone sum rules analysis of the semi-leptonic $D^+_s\to f_0(980)(\to\pi^+\pi^-)e^+\nu_e$ decay incorporating $f_0(980)$ mixing state twist-2 distribution amplitudes}

\author{Dan-Dan Hu}
\email{hudd@stu.cqu.edu.cn}
\author{Xing-Gang Wu}
\email{wuxg@cqu.edu.cn}
\address{Department of Physics, Chongqing Key Laboratory for Strongly Coupled Physics, Chongqing University, Chongqing 401331, P.R. China}
\author{Hai-Jiang Tian}
\address{Department of Physics, Guizhou Minzu University, Guiyang 550025, P.R. China}
\author{Tao Zhong\footnote{Corresponding author}}
\email{zhongtao@gzmu.edu.cn}
\author{Hai-Bing Fu}
\email{fuhb@gzmu.edu.cn}
\address{Department of Physics, Guizhou Minzu University, Guiyang 550025, P.R. China}
\address{Institute of High Energy Physics, Chinese Academy of Sciences, Beijing 100049, P.R.China}

\date{\today}

\begin{abstract}

The isospin-singlet scalar meson $f_0(980)$ is hypothesized to consist of two energy eigenstates, forming a mixture of $\frac{1}{\sqrt{2}}(\bar{u}u + \bar{d}d)$ and $\bar{s}s$. Building on this framework, we apply the QCD sum rules approach in the background field theory to compute the $f_0(980)$ decay constant, yielding $f_{f_0}(\mu_0=1\,\text{GeV}) = 0.386\pm0.009 \, \text{GeV}$. Subsequently, we derive the first two $\xi$-moments of the leading-twist light-cone distribution amplitude $\phi_{2;f_0}$. Using QCD light-cone sum rules, we then calculate the $D^+_s \to f_0(980)$ transition form factor (TFF) $f_+(q^2)$, obtaining $f_+(0) = 0.516^{+0.027}_{-0.024}$ at the large-recoil point. We extend $f_+(q^2)$ to the full physical region via a simplified series expansion parameterization, enabling calculations of the differential decay widths and the branching fraction $\mathcal{B}(D^+_s \to f_0(980)(\to \pi^+\pi^-)e^+\nu_e) = (1.783^{+0.227}_{-0.189}) \times 10^{-3}$. Our theoretical predictions align well with the latest BESIII Collaboration measurements within reasonable errors.

\end{abstract}

\date{\today}


\maketitle

\section{Introduction}

Light scalar mesons present a longstanding classification challenge in particle physics. The understanding of the structure of scalar mesons with masses around $1~{\rm GeV}$ is one of the prominent topics in modern hadronic physics. Originally interpreted as orbitally excited quark - antiquark ($q{\bar q}$) states, these particles have sparked debates over tetraquark ($q{\bar q}q{\bar q}$) or molecular configurations, particularly due to their strong coupling to $\pi\pi$ and $KK$ thresholds which complicate traditional spectroscopic analysis. The ambiguity arises from the interplay between confinement and chiral symmetry breaking, driving investigations into non-perturbative quantum chromodynamics (QCD) effects, multiquark dynamics, and hadronic molecules~\cite{Black:1998wt}. The unique properties of light scalar mesons such as $f_0(500), f_0(980)$ and $a_0(980)$ have been a central focus in hadron physics research for decades. Unraveling their internal quark-gluon substructure is a cornerstone of investigating QCD in the non-perturbative regime, as these systems provide critical insights into the dynamical mechanisms underlying chiral symmetry breaking - a phenomenon fundamental to understanding quark confinement and hadron mass generation in the low-energy sector of QCD.

The weak interaction between leptons and hadrons in the final state renders semi-leptonic decays of charm mesons an ideal experimental framework for probing the quark components in the wave functions of light scalar mesons~\cite{Achasov:2012kk}. This ``clean'' decay system offers dual advantages: first, the spectator light quark mechanism dominates scalar state formation, enabling precise analysis of their quark flavor composition through comparisons of Cabibbo-favored and Cabibbo-suppressed processes; second, extracting the hadronic form factors that characterize strong interactions between final-state quarks provides critical experimental benchmarks for testing non-perturbative theoretical approaches such as the lattice QCD and the QCD sum rules (QCD SR)~\cite{Sekihara:2015iha}. Notably, the form factors and branching fractions of these decays are highly sensitive to the internal structure of light scalar states, making investigations of their dynamics a core experimental tool for unraveling the intrinsic nature of these particles. This endeavor bridges the gap between precision measurements of weak decays and strong interaction theories~\cite{Soni:2020sgn}.

In this paper, we focus on the light scalar meson $f_0(980)$. Its mass and width are measured as $990\pm20~{\rm MeV}$ and $10-100~{\rm MeV}$, according to Ref.~\cite{PDG:2024}. Given its mass below $1~{\rm GeV}$, competing interpretations of its internal quark structure have emerged. Analyses of the $f_0(980)$-meson as a quarkonium ($q\bar{q}$) state were performed in Refs.~\cite{Tornqvist:1995kr, vanBeveren:1998qe, Morgan:1993td, Tornqvist:1995ay, Ke:2009ed}. Different scenarios for the admixture of nonstrange and strange $s\bar{s}$ components have been developed, which range from a pure or dominant $s\bar{s}$ state to mixtures with $n\bar{n}=(u\bar{u}+d\bar{d})/\sqrt{2}$. Alterative models propose composite structures, such as diquark-diantiquark ($q^2\bar{q}^2$)~\cite{Jaffe:1976ig, Alford:2000mm, Fariborz:2009cq}, compact tetraquarks ($q{\bar q}q{\bar q}$)~\cite{Brito:2004tv, Achasov:2010fh, Kim:2017yvd} or meson-meson bound states~\cite{Baru:2003qq, Branz:2007xp, Lee:2013mfa}. Within this mass range, the possibility of a scalar glueball bound state has also been considered~\cite{Robson:1977pm, Narison:1996fm, Lebiedowicz:2020bwo}. Resolving the fundamental properties of this intriguing hadron state therefore requires more precise experimental measurements, advanced non-perturbative QCD techniques, and continued theoretical investigations.

From the experimental perspective, the BABAR collaboration identified evidence for the $f_0(980)$ resonance in semi-leptonic decays by analyzing the $f_0(980)\to K^+K^-$ channel, where interference between an $S$-wave and the dominant $P$-wave $\phi$ decay was observed~\cite{BaBar:2008gpr}. The CLEO collaboration pioneered the first measurement of the product branching fraction for $D^+_s\to f_0(980)$ with the cascade decay $f_0(980)\to\pi^+\pi^-$~\cite{CLEO:2009dyb}. In 2009, they proposed that the $f_0(980)$ wave function combines strange and non-strange quark-antiquark components. Analyzing $D^+_s\to f_0(980)$ decay branching fractions, they determined a mixing angle consistent with $|{\bar s}s\rangle$ dominance, providing evidence for why $f_0(980)$ decays predominantly to $\pi^+\pi^-$ rather than $K^+K^-$~\cite{CLEO:2009ugx}. The BESIII collaboration first measured $D_s^+\to f_0(980)$ in 2019 and observed the $S$-wave contribution, setting an upper limit of $\mathcal{B}(D^+_s\to f_0(980)e^+\nu_e, f_0(980)\to\pi^+\pi^-)<2.8\times 10^{-5}$~\cite{BESIII:2018qmf}. Recently, BESIII collaboration updated these measurements with improved precision, reporting $\mathcal{B}(D^+_s\to f_0(980)e^+\nu_e, f_0(980)\to\pi^0\pi^0)=(0.79\pm0.14\pm0.03)\times 10^{-3}$ for the neutral channel~\cite{BESIII:2021drk} and $\mathcal{B}(D^+_s\to f_0(980)e^+\nu_e, f_0(980)\to\pi^+\pi^-)=(1.72\pm0.13\pm0.10)\times 10^{-3}$ for the charged one~\cite{BESIII:2023wgr}. More notably, the BESIII collaboration extracted the $D_s^+\to f_0(980)$ transition form factor (TFF) under the Flatt\'{e} resonant model with an integrated luminosity of $7.33~{\rm fb^{-1}}$, yielding $f_+(0)=0.518\pm0.018\pm0.036$ at the full recoiled point, including statistical and systematic uncertainties~\cite{BESIII:2023wgr}. Leveraging the Flatt\'{e} formula, BESIII's latest measurements further constrain the $f_0(980)$-meson's properties. Using the relation between the branching fraction (BF) and the mixing angle $\theta$ in the $q{\bar q}$ mixture model $\sin\theta|\frac{1}{\sqrt{2}}({\bar u}u+{\bar d}d)\rangle+\cos\theta|{\bar s}s\rangle$~\cite{Colangelo:2010bg, Shi:2015kha}, they confirmed the dominance of the $s{\bar s}$ component~\cite{BESIII:2023wgr}.

Theoretical predictions for the hadronic TFFs rely on assumptions about the structure of $f_0(980)$, and calculations of the BFs for the semi-leptonic decay $D^+_s\to f_0(980)(\to\pi^+\pi^-)e^+\nu_e$ are highly sensitive to its internal composition. Thus, measurements for these quantities can help unravel its nature. Currently, diverse theoretical groups have predicted the TFFs and the BFs of $D^+_s\to f_0(980)(\to\pi^+\pi^-)e^+\nu_e$, including the light-front quark model (LFQM)~\cite{Ke:2009ed}, chiral unitarity approach ($\chi$UA)~\cite{Sekihara:2015iha}, covariant confined quark model (CCQM)~\cite{Soni:2020sgn}, covariant light-front dynamics (CLFD)~\cite{El-Bennich:2008rkp}, three-point sum rules (3PSR)~\cite{Bediaga:2003zh, Aliev:2007uu, Bediaga:2003hr} and light-cone sum rules (LCSR)~\cite{Cheng:2023knr, Colangelo:2010bg, Shi:2015kha}. Existing theoretical predictions for TFFs and BFs exhibit significant discrepancies with experimental measurements. To systematically address this gap, we will employ the QCD sum rule (QCD SR) approach to develop a comprehensive theoretical description of the $D^+_s\to f_0(980)(\to\pi^+\pi^-)e^+\nu_e$ decay process. This description incorporates the $f_0(980)$ decay constant, twist-2 light-cone distribution amplitude (LCDA) under the LCSR framework, the $D^+_s\to f_0(980)$ TFF and the related BF, aiming to reconcile theoretical predictions with experimental data.

The LCSR approach~\cite{Braun:1988qv, Balitsky:1989ry, Chernyak:1990ag, Ball:1991bs} is a powerful tool for determining non-perturbative parameters of hadronic states. In this framework, the amplitude of the process is factorized into a perturbatively calculable short-distance component and a non-perturbative component encoded in LCDAs. Applying LCSR involves performing an operator product expansion (OPE) in the light-cone vicinity ($x^2 \approx 0$), after which the resulting non-perturbative hadronic matrix elements are then parameterized using LCDAs of increasing twists. The higher twist terms are generally power suppressed, which may also be effectively suppressed within proper Borel windows. In LCSR calculations for the TFFs and BFs of $D^+_s \to f_0(980)(\to \pi^+\pi^-)e^+\nu_e$, the decay constant and twist-2 LCDA of the $f_0(980)$ meson serve as critical input parameters. Despite their fundamental role, theoretical investigations of these observables remain limited. While Refs.\cite{Cheng:2005nb, DeFazio:2001uc} pioneered the determination of $f_0(980)$ decay constants and twist-2 LCDAs using QCD SR approach, this analysis relied on the simplifying assumption of a pure strange-antistrange $\bar{s}s$ configuration for the $f_0(980)$. This approximation overlooks potential contributions from non-valence components or mixing effects with other scalar meson states, which may introduce systematic uncertainties in subsequent phenomenological analyses. In this paper, we model the $f_0(980)$ as a mixed state comprising $\frac{1}{\sqrt{2}}(\bar{u}u + \bar{d}d)$ and $\bar{s}s$ quark components. Using QCD sum rules within the background field theory framework (BFTSR), we systematically compute its decay constant and twist-2 LCDA.

Moreover, the BFTSR approach effectively incorporates the non-perturbative effects and systematically describes vacuum condensates from a field-theoretic perspective. Within this framework, the QCD Background Field Theory (BFT)~\cite{Govaerts:1983ka, Govaerts:1984bk, Huang:1989gv} provides a physical interpretation of both perturbative and non-perturbative QCD properties, establishing a rigorous foundation for deriving well-known Shifman-Vainshtein-Zakharov (SVZ) sum rules~\cite{Shifman:1978bx, Shifman:1978by} relevant to hadron phenomenology. Consequently, BFT is well-suited for computing meson LCDAs. At present, the BFTSR approach has been successfully applied to calculate LCDAs of multiple mesons, enhancing the utility of QCD sum rules in hadron physics. This approach not only yields more reliable theoretical predictions for experimental observations but also provides a powerful tool for probing the non-perturbative mechanisms and hadron structure in QCD. Subsequently, we employ the LCSR method to compute the $D^+_s \to f_0(980)$ TFFs. By incorporating the decay width formula recently reported by the BESIII Collaboration, we derive the corresponding differential decay widths and BFs. Our results substantially reduce the discrepancy between theoretical predictions and experimental measurements, offering new theoretical constraints for unraveling the structure of light scalar mesons.

This paper is structured as follows. In Sec.~\ref{sec:2}, we first briefly introduce the decay width for the semileptonic $D^+_s \to f_0(980)(\to \pi^+\pi^-)e^+\nu_e$. We then outline the principles of the BFTSR framework for calculating the moments of the $f_0(980)$ meson's twist-2 LCDA, as well as the LCSR framework for calculating the TFF associated with this semileptonic decay. In Sec.~\ref{sec:3}, we present our numerical results and conduct a detailed comparison with other experimental and theoretical predictions. Section~\ref{sec:summary} is dedicated to a summary.

\section{Calculation Technology}\label{sec:2}

\begin{figure}[htb!]
\begin{center}
\includegraphics[width=0.45\textwidth]{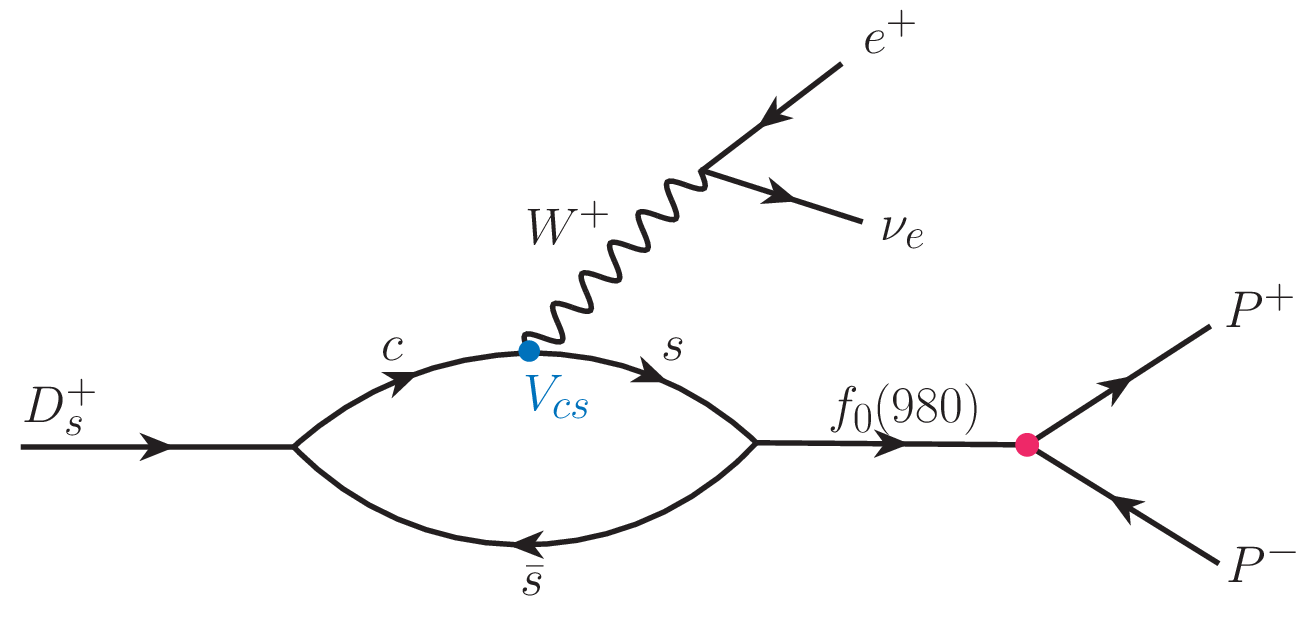}
\end{center}
\caption{The $D^+_s\to f_0(980)(\to P^+ P^-)e^+\nu_e$ decay diagram, where $P$ represents $\pi$ or $K$, respectively.} \label{fig:fm}
\end{figure}

The decay diagram at hadronic level for the process $D^+_s\to f_0(980)(\to P^+ P^-)e^+\nu_e$ is shown in Fig.~\ref{fig:fm}, where $P$ represents $\pi$ or $K$, respectively. After emitting the off-well $W^+$-boson, the hadronic sector is the ${\bar s}s$ which will couple to the isosinglet component $f_0(980)$. At the energy scale near $1~{\rm GeV}$, the $f_0(980)$-meson, due to its inherent instability, exhibits significant coupling with the $\pi^+\pi^-$ through strong interaction~\cite{CLEO:2009ugx}. Then the decay amplitudes for the $D^+_s\to f_0(980)e^+\nu_e\equiv\pi^+\pi^-e^+\nu_e$ is given as~\cite{Wang:2016wpc}
\begin{align}
{\mathcal A}(D^+_s\to\pi^+\pi^-e^+\nu)&=\hat{A}\bigg( \frac{i}{D_{f_0}}\times ig_{\pi^+\pi^-} \bigg),
\end{align}
where $D_{f_0}$ is the $f_0(980)$-meson denominators of the resummed propagators, encompasses the self-energy correction, the width effect, and the coupling information of the $f_0(980)$. It is typically expressed in the form:
\begin{align}
D_{f_0}(s)=s-m^2_{f_0}+\Pi_{f_0}(s),
\end{align}
where $\Pi_{f_0}(s)$ is self-energy function. The real part of the self-energy function causes a shift in the propagator pole, reflecting the correction of the $f_0(980)$-meson mass from the bare mass $m_{f_0}$ to the physical mass $s$. The imaginary part is directly related to the decay width $\Gamma_{f_0}$, embodying the probability of $f_0(980)$ decaying into $\pi^+\pi^-$ or $K^+K^-$ via strong interaction. Consequently, the expression for $D_{f_0}$ can be further articulated as~\cite{Flatte:1976xv, LHCb:2014ooi}:
\begin{align}
D_{f_0}(s)=s-m^2_{f_0}+i{\sqrt s}\left[\Gamma^{f_0}_{\pi^+\pi^-}(s)+\Gamma^{f_0}_{K^+K^-}(s)\right].
\end{align}
$\Gamma^{f_0}_{\pi^+\pi^-}(s)$ and $\Gamma^{f_0}_{K^+K^-}(s)$ are nonzero mass-dependent decay width, which the specific expressions are respectively:
\begin{align}
&\Gamma^{f_0}_{\pi^+\pi^-}(s)=\frac{1}{16\pi\sqrt{ s}}|g_{\pi^+\pi^-}|^2\rho_{\pi\pi}(s),
\nonumber\\
&\Gamma^{f_0}_{K^+K^-}(s)=\frac{1}{16\pi\sqrt{s}}|g_{ K^+K^-}|^2\rho_{K{\bar K}}(s),
\end{align}
where $g_{\pi^+\pi^-}=\sqrt{16\pi g_1}$ and $g_{K^+K^-}=\sqrt{16\pi g_2}$, with the constants $g_1$ and $g_2$ are the $f_0(980)$ couplings to $\pi^+\pi^-$ and $K^+K^-$ final states, respectively. The factors $\rho_{\pi\pi}$ and $\rho_{K{\bar K}}$ are given by Lorentz invariant phase space~\cite{Achasov:2020qfx}
\begin{align}
&\rho_{\pi\pi}(s)=\sqrt{\left[1-\frac{(m_\pi-m_\pi)^2}{s}\right]\left[1-\frac{(m_\pi+m_\pi)^2}{s}\right]},
\\
&\rho_{K{\bar K}}(s)\!=\!\sqrt{\left[1\!-\!\frac{(m_K-m_{\bar K})^2}{s}\right]\!\!\left[1\!-\!\frac{(m_K+m_{\bar K})^2}{s}\right]}.
\end{align}
The amplitude $\hat{A}$ can be expressed in terms of the TFFs~\cite{Sun:2010nv}:
\begin{align}
\langle f_0(980)(p)&|\bar{s} \gamma_\mu \gamma_5 c|D^+_s(p+q)\rangle
\nonumber\\
&= -2if_+(q^2)p_\mu- i[f_+(q^2) + f_-(q^2)] q_\mu,
\label{Eq:me}
\end{align}
where $f_{\pm}(q^2)$ are TFFs. Since this transition is between a pseudoscalar $D^+_s$ and a scalar $f_0(980)$, only the axial-vector part of the current contribution. The term containing $f_-(q^2)$ gets mulitiplied by the positron mass squared, and becomes vanishingly small due to chiral suppression~\cite{Artuso:2008vf}. The double differential decay width is then derived as~\cite{BESIII:2023wgr, Achasov:2020qfx}:
\begin{align}
&\frac{d\Gamma(D^+_s\to f_0(980)(\to\pi^+\pi^-)e^+\nu_e)}{ds\ dq^2}
\nonumber\\
& =\frac{G^2_F|V_{cs}|^2}{192\pi^3m^3_{D^+_s}}\lambda^{3/2}(m^2_{D^+_s},m^2_{f_0},q^2)|f_+(q^2)|^2\frac{P(s)}{\pi},
\label{Eq:dgamma}
\end{align}
where $s$ is the invariant mass square of the two pseudo-scalars $\pi^+, \pi^-$. For the $f_0(980)$, one may follow the BESIII analysis of the $J/\Psi\to\phi\pi^+\pi^-$~\cite{BES:2004twe}, $s_{\rm min}=0.9~{\rm GeV}/c^2, s_{\rm max}=1.0~{\rm GeV}/c^2$. Due to the limitation of the physical region $q^2$, in the actual calculation process, our value is $s_{\rm min}\leq s \leq (m_{D^+_s} - m_{f_0})^2$. $G_F$ is the Fermi constant, $|V_{cs}|$ is the Cabibbo-Kobayashi-Maskawa (CKM) matrix element, $\lambda(x,y,z)=x^2+y^2+z^2 - 2xy - 2xz - 2yz$, and $P(s)$ is based on the relativistic Flatt\'{e} formula due to the open $K^+K^ - $ channel as follows:
\begin{align}
\hspace{-0.2cm} P(s)=\frac{g_1\rho_{\pi\pi}}{|m^2_{f_0} \!-\! s \!-\! i[g_1 \rho_{\pi\pi} (s) \!+\! g_2\rho_{K{\bar K}}(s)]|^2}.
\end{align}
Here the $\rho_{\pi\pi}$ and $\rho_{K{\bar K}}$ are individual phase space factors.

The TFF $f_+(q^2)$ can be calculated by using the LCSR method. In order to uniformly express the derivation process of the
LCSRs for $D^+_s\to f_0(980)$ TFFs, we introduce the following vacuum-to-$f_0(980)$ correlator
\begin{align}
\Pi_{\mu}(p,q) =&i\int d^4xe^{iq\cdot x}\langle f_0(980)(p)|T\{{\bar s}(x)\gamma_\mu\gamma_5c(x),
\nonumber\\
&\times{\bar c}(0)i\gamma_5 s(0)\}|0\rangle
\nonumber\\
&=F[q^2,(p+q)^2]p_\mu+\widetilde{F}[p^2,(p+q)^2]q_\mu .
\label{Eq:1}
\end{align}
We first calculate the correlator \eqref{Eq:1} in QCD. Under the kinematic conditions $q^2\ll m^2_c$ and $(p+q)^2\ll m^2_c$ (where $m_c$ denotes the heavy charm quark mass), the heavy quark propagating within the correlation function exhibits significant virtuality. This regime corresponds to spatial separations approaching the light-cone limit $x^2\to 0$~\cite{Duplancic:2008ix}, as dictated by the dominance of small momentum transfer scales relative to the heavy quark mass. In this scenario, the OPE framework simplifies through the contraction of heavy quark fields via Wick's theorem, enabling light-cone expansion of the heavy quark propagator. The resulting formalism allows for the factorization of short-distance perturbative effects from long-distance non-perturbative dynamics, characteristic of LCDA analyses. Thus one can contract the heavy quark fields, and the following light-cone expansion of the heavy quark propagator
\begin{align}
\hspace{-0.3cm}\langle 0| c^i_\alpha(x){\bar c}^j_\beta(0)|0\rangle\!=\!i\!\int\!\frac{d^4k}{(2\pi^4)}e^{ - ik\cdot x}\bigg(\delta^{ij}\frac{\DS k+m_c}{k^2 - m^2_c} + \cdots \bigg)_{\alpha\beta}.
\end{align}
Then the vacuum-to-$f_0(980)$ matrix element can be expanded in terms of the $f_0(980)$-meson LCDA's of increasing twist, {\it i.e.}
\begin{align}
\langle f_0&(980)(p)|{\bar s}^i_\alpha(x)s^j_\beta(0)|0\rangle
\nonumber\\
&=\frac{\delta^{ji}}{12}{\bar f}_{f_0}\int^1_0 due^{iup\cdot x}\bigg[\DS p\phi_{2;f_0}(u) + m_{f_0}\phi^p_{3;f_0}(u)
\nonumber\\
&-  \frac{1}{6}m_{f_0} \sigma_{\mu\nu} p^{\mu} x^{\nu} \phi^\sigma_{3;f_0}(u)\bigg]_{\beta\alpha} +\cdots
\end{align}
with the $\phi_{2;f_0}(u)$ and $\phi^{p,\sigma}_{3;f_0}(u)$ are the two-particle twist-2, 3 LCDA respectively. Using the OPE for the invariant amplitudes $F$ and $\widetilde{F}$, we obtain
\begin{align}
F_{\rm QCD}[q^2,(p+q)^2]&=i m_c {\bar f}_{f_0}\int^1_0\frac{du}{m_c^2 - (up+q)^2}
\nonumber\\
&\times\bigg\{ \phi_{2;f_0}(u)  -  \frac{m_{f_0}}{m_c} u \phi^p_{3;f_0}(u)  - \frac{m_{f_0}}{6m_c}
\nonumber\\
&\times \bigg[2+\frac{m^2_c+q^2 - u^2p^2} {m^2_c - (up+q)^2} \bigg] \phi^\sigma_{3;f_0}(u) \bigg\},
\\
\widetilde{F}_{\rm QCD}[q^2,(p+q)^2] &=i \, m_c \, {\bar f}_{f_0} \int^1_0 \frac{du}{m_c^2 \,-\, (up \,+ \,q)^2}
\nonumber\\
&\times\bigg\{\,  - \, \frac{m_{f_0}}{m_c} \, \phi^\sigma_{3;f_0}(u) \,  - \,  \frac{m_{f_0}}{6 m_c}\ \bigg[\ 1
\nonumber\\
&- \frac{m^2_c+q^2 - u^2p^2} {m^2_c - (up+q)^2}\bigg] \frac{\phi^\sigma_{3;f_0}(u)}{u}
\bigg\}.
\end{align}

On the other hand, the correlation function \eqref{Eq:1} in the timelike $q^2$-region can be treated with the hadronic representation, which is achieved by inserting a complete set of intermediate states with the same quantum numbers in the correlation function. By isolating the pole term of the lowest pseudoscalar $D^+_s$-meson, we obtain the following representation of the correlation function from the hadronic side:
\begin{align}
&\Pi_\mu(p,q)
\nonumber\\
&=\frac{\langle f_0(980)(p)|{\bar s}\gamma_\mu\gamma_5c|D^+_s(p+q)\rangle\langle D^+_s(p+q)|{\bar c}i\gamma_5 s|0\rangle}{m^2_{D^+_s} - (p+q)^2}
\nonumber\\
&+\sum_{H}\frac{\langle f_0(980)(p)|{\bar s}\gamma_\mu\gamma_5c|D^H_s(p+q)\rangle\langle D^H_s(p+q)|{\bar c}i\gamma_5 s|0\rangle}{m^2_{D^H_s}-(p+q)^2}
\end{align}
where $\langle D^+_s(p+q)|{\bar c}i\gamma_5 s|0\rangle=m^2_{D^+_s}f_{D_s^+}/(m_c+m_s)$ with $f_{D_s^+}$ is $D^+_s$-meson decay constant. By substituting the parameterized Eq.~\eqref{Eq:me} (as the TFF) into the above equation, the invariant amplitude in the hadronic representation can be obtained as follows:
\begin{align}
F_{\rm had}[p^2,(p+q)^2]&=\frac{-2im^2_{D^+_s}f_{D_s^+}f_+(q^2)}{(m_c+m_s)[m^2_{D^+_s}-(p+q)^2]}+\cdots
\nonumber\\
\widetilde{F}_{\rm had}[p^2,(p+q)^2]&=\frac{-im^2_{D^+_s}f_{D_s^+}[f_+(q^2)+f_-(q^2)]}{(m_c+m_s)[m^2_{D^+_s}-(p+q)^2]}+\cdots
\end{align}
Eq.~\eqref{Eq:dgamma} indicates that the decay width $\Gamma(D^+_s\to f_0(980)(\to\pi^+\pi^-)e^+\nu_e)$ primarily depends on the TFF $f_+(q^2)$. Therefore, we will explain its derivation in some more detail in the following. One can derive a general dispersion relation for $F[q^2,(p+q)^2]$ and subsequently apply the usual Borel transformation with respect to the momentum squared $(p+q)^2$ of the heavy meson~\cite{Belyaev:1993wp} to suppress the contributions from
the unknown continuum states and higher resonances:
\begin{align}
F(q^2,M^2)=\int^\infty_{t_{\rm min}}\rho(q^2,s)e^{-s/M^2}ds
\label{Eq:2}
\end{align}
with $t_{\rm min}=(m_c+m_s)^2$. Here the spectral density is given by
\begin{align}
\rho(q^2,s)=&\frac{1}{\pi}{\rm Im}F_{\rm had}(q^2,s)
\nonumber\\
&=\delta(s-m^2_{D^+_s})\frac{-2im^2_{D^+_s}f_{D_s^+}}{m_c+m_s}f_+(q^2)
\nonumber\\
&~~+\frac{1}{\pi}{\rm Im}F_{\rm QCD}(q^2,s)\theta(s-s_0).
\label{Eq:3}
\end{align}
In this formulation, the contributions from excited states and continuum states contributions in $F_{\rm had}(q^2,s)$ are incorporated through the imaginary part of the QCD correlation function ${\rm Im}F_{\rm QCD}(q^2,s)/\pi$, with non-perturbative effects demarcated by an effective continuum threshold parameter $s_0$ under the quark-hadronic duality approximation~\cite{Shifman:1978bx, Shifman:1978by}. The Borel transformation, as implemented in Eq.~\eqref{Eq:2}, systematically suppresses potential subtraction arising from dispersion relation while enhancing the convergence of the OPE by exponentially restraining excited states. Substituting Eq.~\eqref{Eq:3} into Eq.~\eqref{Eq:2}, one can get
\begin{align}
F(q^2,M^2)&=\frac{-2im^2_{D^+_s}f_{D_s^+}f_+(q^2)}{m_c+m_s}e^{m^2_{D^+_s}/M^2}
\nonumber\\
&+\frac{1}{\pi}\int_{s_0}^{\infty}F_{\rm had}(q^2,s)e^{-s/M^2}ds
\label{Eq:4}
\end{align}
On the other hand, the invariant amplitude after Borel transformation can also be written as
\begin{align}
F(q^2,M^2)=\frac{1}{\pi}\int_{t_{\rm min}}^{\infty}F_{\rm had}(q^2,s)e^{-s/M^2}ds
\label{Eq:5}
\end{align}
Equating Eqs.~\eqref{Eq:4} and \eqref{Eq:5}, the LCSR of TFF $f_+(q^2)$ can be obtained as
\begin{widetext}
\begin{eqnarray}
f_+(q^2) &=& \frac{(m_c+m_s) \bar{f}_{f_0}}{2m_{D_s}^2 f_{D_s}} e^{m^2_{D^+_s}/M^2} \bigg\{ \int^1_{u_0} du e^{-(m_c^2-\bar{u}q^2+u\bar{u}m_{f_0}^2)/(uM^2)}
\bigg[-m_c \frac{\phi_{2;f_0}(u)}{u}
\nonumber\\
& & + m_{f_0} \bigg(\frac{2}{u} + \frac{4u m_c^2 m_{f_0}^2}{(m_c^2-q^2+u^2m_{f_0}^2)^2} - \frac{m_c^2+q^2-u^2m_{f_0}^2}{m_c^2-q^2+u^2m_{f_0}^2} \frac{d}{du} \bigg)\frac{\phi_{3;f_0}^\sigma(u)}{6}+ m_{f_0}
\nonumber\\
&& \times \phi_{3;f_0}^p(u) \bigg] +e^{-(m_c^2 - \bar{u}q^2 + u\bar{u}m_{f_0}^2)/(uM^2)} \frac{m_{f_0}}{6m_c}\frac{m_c^2 + q^2 - u^2m_{f_0}^2}{m_c^2 - q^2 + u^2m_{f_0}^2} \phi_{3;f_0}^\sigma(u) \Big|_{u\to1}\bigg\}.
\label{Eq:fp}
\end{eqnarray}
\end{widetext}
with $u_0=[\sqrt{(q^2-s_0+m^2_{f_0})^2+4m^2_{f_0}(m^2_c-q^2)}+q^2-s_0+m^2_{f_0}]/2m^2_{f_0}$. Based on the conformal spin invariance, the LCDAs can be expanded in terms of Gegenbauer polynomials $C^{3/2}_n$. For the twits-3 LCDAs, we use the asymptotic form, {\it i.e.} the first term of the Gegenbauer  expansion~\cite{Colangelo:2010bg},
\begin{align}
&\phi_{3;f_0}^p(u)=1,
\nonumber\\
&\phi_{3;f_0}^\sigma(u)=6u(1-u).
\end{align}

According to Eq.~\eqref{Eq:dgamma}, it can be found that the width of $D^+_s\to f_0(980)$ is mainly $\pi^+\pi^-$, which also indicates that $f_0(980)$ is not purely $s{\bar s}$ state, but should contains the composition of $u{\bar u}$ and $d{\bar d}$ pairs. Under ${\rm SU}_f(3)$ symmetry, the light scalar meson $f_0(980)$ can have two possible quark structures, leading to two possible mixing scenarios. For the ${\bar q}^2q^2$ structure, the $f_0(980)$ can be expressed as follows~\cite{Maiani:2004uc, Kim:2018zob}
\begin{align}
\!\!\! |f_0(980)\rangle=|\frac{1}{\sqrt{2}}(u{\bar u}+d{\bar d})s{\bar s}\rangle\cos\theta_I+|u{\bar u}d{\bar d}\rangle\sin\theta_I,
\end{align}
It has been shown that $\theta_I=174.6^\circ$~\cite{Maiani:2004uc}. The light scalar meson $f_0(980)$ have another possible ${\bar q}q$ structure~\cite{Aliev:2007uu, Colangelo:2010bg, Shi:2015kha, Ke:2009ed}
\begin{align}
|f_0(980)\rangle=|s{\bar s}\rangle\cos\theta+|n{\bar n}\rangle\sin\theta,
\end{align}
with $\theta$ is mixing angle and $n{\bar n}=(u{\bar u}+d{\bar d})/\sqrt{2}$. Experimental implications for the mixing angle have been discussed in detail in~\cite{Alford:2000mm, Cheng:2002ai, Gokalp:2004ny}. Here, we primarily adopt the results obtained by the BESIII'24 Collaboration, namely: $\theta=(19.7\pm12.8)^\circ$~\cite{BESIII:2023wgr}. In general, the $f_0(980)$-meson twist-2 LCDA $\phi_{2;f_0}(u)$ can be expanded as a Gegenbauer polynomial series~\cite{Cheng:2005nb, Braun:2003rp}
\begin{align}
\phi_{2;f_0}(x,\mu)&={\bar f}_{f_0}(\mu)6x{\bar x}\left[ B_0(\mu)+\sum^\infty_{n=1}B_n(\mu)C^{3/2}_n(\xi)\right]
\label{Eq:da}
\end{align}
where ${\bar x}=(1-x)$, $\xi=(2x-1)$ and $B_0, B_n$ are Gegenbauer moments, $C^{3/2}_n$ are the Gegenbauer polynomials. $f_{f_0}$ is the $f_0(980)$-meson decay constant and ${\bar f}_{f_0}$ indicates that the decay constant $f_{f_0}$ varies with the energy scale. Because of the charge conjugation invariance, all even Gegenbauer moments of $\phi_{2;f_0}(x)$ vanish, so that $B_{2m}=0$ for $m=(0,1,\cdots)$ in Eq.~\eqref{Eq:da}. As for the odd moments, we will adopt BFTSR-calculated results.

According to the $f_0(980)$ quark components, its twist-2 LCDA can be decomposed into two parts for calculation, namely:
\begin{align}
\phi_{2;f_0}(x,\mu)=\cos\theta\phi^{(s\bar s)}_{2;f_0}(x,\mu)+\sin\theta\phi^{(n\bar{n})}_{2;f_0}(x,\mu).
\label{Eq:phimix}
\end{align}

Let's first introduce the calculation procedures for the $\xi$-moments of $\phi^{(s\bar s)}_{2;f_0}(x,\mu)$ for the first part. The twist-2 LCDA $\phi^{(s\bar s)}_{2;f_0}(x,\mu)$ and the twist-3 LCDA $\phi^{p(s\bar s)}_{3;f_0}(x,\mu)$ for the scalar meson $f_0(980)$ made of the $s{\bar s}$ component are given by~\cite{Cheng:2005nb}
\begin{align}
& \langle 0|{\bar s}(z)\gamma_\mu s(-z)|f_0(980)\rangle =p_\mu{\bar f}^{(s\bar s)}_{f_0}\int^1_0dxe^{i\xi (p\cdot x)}\phi^{(s\bar s)}_{2;f_0}(x,\mu)
\nonumber\\
& \langle 0|{\bar s}(z)s(-z)|f_0(980)\rangle = m_{f_0}{\bar f}^{(s\bar s)}_{f_0}\int^1_0dxe^{i\xi (p\cdot x)}\phi^{p(s\bar s)}_{3;f_0}(x,\mu)
\label{Eq:phi}
\end{align}
where $z^2=0$ and $m_{f_0}$ is $f_0(980)$-meson mass. Expanding both sides of Eq.~\eqref{Eq:phi} into $z$-series, one gets
\begin{align}
& \langle 0|{\bar s}(0) \DS z {(iz \cdot \tensor D )^n} s(0)|f_0(980)\rangle = {\bar f}^{(s\bar s)}_{f_0}(z\cdot p)^{n+1}\langle\xi^{(s\bar s),n}_{2;f_0}\rangle
\nonumber\\
& \langle 0|{\bar s}(0)s(0)|f_0(980)\rangle =m_{f_0}{\bar f}^{(s\bar s)}_{f_0}\langle\xi^{p(s\bar s),0}_{3;f_0}\rangle
\end{align}
The covariant derivative satisfies the relation $(iz \cdot \tensor D )^n=(iz\cdot\mathop{D}\limits^\rightarrow-iz\cdot \mathop{D}\limits^\leftarrow)$. Based on the basic procedure of QCD SR, one can adopt the following two-point correlator to derive the sum rules for $f_0(980)$-meson twist-2 LCDA moments $\langle\xi^{(s\bar s),n}_{2;f_0}\rangle|_\mu$, which can be read off
\begin{align}
\Pi^{(s\bar s)}_{2;f_0}(z,q)&=i\int d^4xe^{iq\cdot x}\langle 0|T\{J^{(s\bar s)}_n(x),\hat{J}_0^{(s\bar s),\dagger}(0)\}|0\rangle
\nonumber\\
&=(z\cdot q)^{n+1}I^{(s\bar s)}_{2;f_0}(q^2)
\label{Eq:cf}
\end{align}
where $J^{(s\bar s)}_n(x)={\bar s}(x) \DS z {(iz \cdot \tensor D )^n} s(x)$ and $\hat{J}_0^{(s\bar s),\dagger}(0)={\bar s}(0)s(0)$ are interpolating currents. $\langle\xi^{(s\bar s),n}_{2;f_0}\rangle|_\mu=\int^1_0dx(2x-1)^n\phi^{(s\bar s)}_{2;f_0}(x,\mu)$ is the $n_{\rm th}$ $\xi$-moment. Because of the G-parity, only odd moments of $\langle\xi^{(s\bar s),n}_{2;f_0}\rangle|_\mu$ are non-zero, {\it i.e.} $n=(1,3,\cdot\cdot\cdot)$. Within the framework of BFTSR, the application of Feynman diagrammatic rules enables a systematic OPE treatment of the correlation function \eqref{Eq:cf} in deep Euclidean region $q^2\ll 0$. This expansion decomposes the correlation function into three distinct components: (1) the $s$-quark propagators $S^s_F(0,x)$ and $S^s_F(x,0)$, representing quark propagation from space-time points `$0$ to $x$' and `$x$ to $0$' respectively; (2) vertex operators $(iz \cdot \tensor D )^n$ are covariant derivative interactions; (3) a series of local gauge-invariant operators of increasing dimensionality. For the Lorentz invariant function $\Pi^{(s\bar s)}_{2;f_0}(z,q)$, the OPE generates an asymptotic series organized by operator dimension, with the leading contributions arising from low-dimensional condensates while higher-dimensional terms are suppressed by powers of the renormalization scale. The expectation values of these operators in the non-perturbative (physical) vacuum are known as vacuum condensates, whose detailed expression can be found in Refs.~\cite{Hu:2021zmy, Zhong:2021epq}. By inserting a complete set of hadronic states into correlator \eqref{Eq:cf} in physical region, whose hadronic representation can be read as
\begin{align}
{\rm Im}I^{\rm had}_{2;f_0}(s)&=\pi m_{f_0}\delta(s-m^2_{f_0}){\bar f} ^{(s\bar s)2} _{f_0} \langle \xi^{(s\bar s),n} _{2;f_0} \rangle| _\mu \langle \xi^{p(s\bar s),0} _{3;f_0} \rangle|_\mu
\nonumber\\
&+{\rm Im}I^{\rm pert}_{2;f_0}(s)\theta(s-s_{f_0})
\end{align}
where $s_{f_0}$ is the continuum threshold, $\langle\xi^{p(s\bar s),0}_{3;f_0}\rangle|_\mu$ is the $O_{\rm th}$ $\xi$-moment of $f_0(980)$ two-particle twist-3 LCDA. Substituting the resulted OPE and hadronic representation of correlator Eq.~\eqref{Eq:cf}, into the following dispersion relation after Borel transformation,
\begin{align}
\frac{1}{\pi}\frac{1}{M^2}\int^{s_{f_0}}_{4m^2_s}ds e^{-s/M^2}{\rm Im}I^{\rm had}_{2;f_0}(s)=L_{M^2} I^{\rm QCD}_{2;f_0}(q^2)
\end{align}
And the final sum rules for the moment of $f_0(980)$-meson twist-2 LCDA is given by
\begin{widetext}
\begin{align}
&\frac{m_{f_0}\bar{f}_{f_0}^{2(s\bar{s})}}{M^2e^{m_{f_0}^{2}/M^2}}\langle \xi _{3;f_0}^{p(s\bar{s}),0}\rangle \langle \xi _{2;f_0}^{(s\bar{s}),n}\rangle =[1 - ( - 1)^n]\biggl\{  - \frac{1}{\pi M^2}\int_{4m_{s}^{2}}^{s_{f_0}}{dse^{ - s/M^2}} \frac{3m_s}{16\pi (n + 1)(n + 2)}
\nonumber\\
&\hspace{1.2cm} \times [2n + 3 + ( - 1)^n] + \left( \frac{1}{M^2}\, + \,\frac{nm_{s}^{2}}{M^4} \right) \langle \bar{s}s\rangle \, + \,\left(  - \,\frac{n}{2M^4}\, + \,\frac{5nm_{s}^{2}}{18M^6} \right) \langle g_s\bar{s}\sigma TGs\rangle
\nonumber\\
&\hspace{1.2cm}  + \frac{2m_s}{27M^6}\langle g_s\bar{s}s\rangle ^2 + L_{M^2}I_{\langle G^2\rangle} + L_{M^2}I_{\langle G^3\rangle} + L_{M^2}I_{\langle q^4\rangle}\bigg\},
\label{Eq:twist2s}
\\
&L_{M^2}I_{\langle G^2\rangle}= - m_s\frac{\langle \alpha _sG^2\rangle}{48\pi M^4}\bigg\{  4\biggr[\ln \biggr( \frac{M^2}{\mu ^2}\!\biggl)\,+\,1-\gamma_E\biggl] \, +\, 2\,\,\frac{n + 2}{n + 1} \, -\, 4(\psi ^{(0)}(n + 1) + \gamma _E)
\nonumber\\
&\hspace{1.2cm} + \theta (n - 2) \biggl[ (6n + 4)\psi _3(n)  + \frac{4n^3 + 10n^2 - 6n - 8}{n(n + 1)}\bigg] + \theta (n - 1)\bigg[ - \frac{2}{n + 1} [ 3n^2
\nonumber\\
&\hspace{1.2cm}  + 5n  + 3+ (n + 1)\psi _2(n)]  - 8(n + 1)\bigg[\psi ^{(0)}(n + 1) + 2\gamma _E - \ln \bigg( \frac{M^2}{\mu ^2} \bigg)-1 \bigg]\bigg]\bigg\},
\label{Eq:Igg} \\
&L_{M^2} I_{\langle G^3\rangle}=m_s\frac{\langle g^3_sfG^3\rangle}{23040\pi^2 M^6}\bigg\{(1584 -7776n)\bigg[ \ln\bigg(\frac{M^2}{\mu^2} \bigg)+\frac{3}{2}-\gamma_E\bigg]  +\bigg[72(108n - 22)
\nonumber\\
&\hspace{1.2cm}
 -96(81n\,-\, 5)\bigg](\psi^{(0)}(\,n  +\!  1) + \gamma_E)  \, -  \frac{12( - 1458n^2  -  1471n  -  121)} {n + 1}+\!  \theta(n\!  - \! 2)
\nonumber\\
&\hspace{1.1cm}
\times\bigg[72(n - 1)\frac{ - 51n^3 - 35n^2  + 80n + 52} {n (n + 1)}  +   144(9n + 13) \psi_3(n) +  864n(n - 1)
\nonumber\\
&\hspace{1.2cm}
\times\bigg[\ln\bigg(\!\frac{M^2}{\mu^2}\! \bigg)\,+\,\frac{3}{2}\,-\,\gamma_E\bigg]\bigg]\,+\, \theta(n\!  - \! 1)\bigg[288(5n + 8)\bigg[ \ln\bigg(\!\frac{M^2}{\mu^2}\! \bigg)\,+\,\frac{3}{2}\,-\gamma_E\bigg]
\nonumber\\
&\hspace{1.2cm}
-  (1440n \,+\, 1824)(\,\psi^{(0)}(n + 1)\,+\,  \gamma_E\,)  \,-  \! \frac{12}{n+1}\,(\,431n^2 \, -\,1281n \, -\, 1030\,)
\nonumber\\
&\hspace{1.2cm}
 -96\psi_2(n)\bigg]\,+\, \theta(n - 3)\bigg[\frac{ 54}{n(n^2 - 1)} (\, - 77n^5 - 249n^4+ 67n^3  -  95n^2  -  294n
\nonumber\\
&\hspace{1.2cm}
-  72)   \,- \, 216\,(20n^2 + 16n + 9)\psi_1(n)\,  -\,  648(n^2 \,+ 5n \,+ 6)\,\bigg[\ln\bigg(\frac{M^2}{\mu^2}\bigg)+\frac{3}{2}
\nonumber\\
&\hspace{1.2cm}
-\gamma_E\bigg] + 648(n^2 + 5n +6)(\psi^{(0)}(n + 1) + \gamma_E)\bigg] \bigg\},
\label{Eq:Iggg}  \\
& L_{M^2} I_{\langle q^4\rangle}=m_s\frac{(2 \,+ \kappa^2) \, \langle g^2_sq{\bar q}\rangle^2}{233280\pi^2 M^6} \,  \bigg\{648(66n - 23)\,\bigg[\ln\bigg(\frac{M^2}{\mu^2}\bigg)\,+\,\frac{3}{2}\,-\,\gamma_E\bigg] \,+ \,14904
\nonumber\\
&\hspace{1.2cm}
 \times (\psi^{(0)}(n + 1) + \gamma_E)\,+ \,\frac{216(27n^2 + 40n + 82)}{n + 1}\,+\, \theta(n - 3)\bigg[\,3564(n^2 + 5n
\nonumber\\
&\hspace{1.2cm}
 + 6)\bigg[\ln\bigg(\frac{M^2}{\mu^2}\bigg)+\frac{3}{2}-\gamma_E\bigg] + 1188(20n^2  +  16n  +  9)\psi_1(n)\, -    3564(n^2  +  5n
\nonumber\\
&\hspace{1.2cm}
+ \, 6)(\,\psi^{(0)} (n  +  1) \,+\,\gamma_E\,)\,+\,\frac{54 }{n(n^2 -  1)}\,(581n^5 \,+ 2355n^4 \,+ 1013n^3  \,-991n^2
\nonumber\\
&\hspace{1.0cm}
  +606n+ 396)\bigg] + \theta(n - 2)\bigg[ - 4752n(n - 1)\bigg[\ln\bigg(\frac{M^2}{\mu^2}\bigg)+\frac{3}{2}-\gamma_E\bigg] +  72(46n
\nonumber\\
&\hspace{1.2cm}
-  45)\psi_3(n) - 108(n - 1)\frac{- 82n^3 - 35n^2 + 83n + 60}{n^2(n + 1)}\bigg] +\theta(n - 1)\,\bigg\{\, 414\,\psi_2(n)
\nonumber\\
&\hspace{1.2cm}
 - 648(3n - 10)\bigg[(\psi^{(0)}(n + 1) + \gamma_E)\bigg] \,+ 648(3n - 10) \bigg[\ln\bigg(\frac{M^2}{\mu^2}\bigg)\,+\frac{3}{2}-\gamma_E\bigg]
\nonumber\\
&\hspace{1.2cm}
 - \frac{18}{n + 1}(1555n^2 + 2283n + 368)\bigg\} \bigg\}.
\label{Eq:Igq44}
\end{align}
In particular, the sum rule \eqref{Eq:twist2s} is regarded as that for $\langle\xi^{p(s\bar s),0}_{3;f_0}\rangle \langle\xi^{(s\bar s),n}_{2;f_0}\rangle$ instead of $\langle\xi^{(s\bar s),n}_{2;f_0}\rangle$ in this work due to dependence of $\langle\xi^{p(s\bar s),n}_{3;f_0}\rangle$ on the Borel parameter as suggested. This assumption can be confirmed by the sum rule of $\langle\xi^{p(s\bar s),n}_{3;f_0}\rangle$ derived form the correlator $i\int d^4xe^{i q\cdot x}\langle 0|\hat{J}^{(s\bar s)}_0(0)\hat{J}^{(s\bar s)\dagger}_0(0)|0\rangle$. Following the above sum rule calculation procedure performed for correlator \eqref{Eq:cf}, we then obtain
\begin{align}
&\frac{m_{f_0}^{2}\bar{f}_{f_0}^{2(s\bar{s})}}{M^2e^{m_{f_0}^{2}/M^2}}\langle \xi _{3;f_0}^{p(s\bar{s}),0}\rangle ^2  =\frac{1}{\pi M^2} \! \int_{4m_{s}^{2}}^{s_{f_0}}{dse^{-s/M^2} \mathrm{Im} I_{3;f_0}^{\mathrm{pert}}(s)}  \! - \! \frac{3 m_s}{M^2}\langle \bar{s}s\rangle \! + \!  \frac{2}{27M^4}\bigg( 4+\frac{5m_{s}^{2}}{M^2} \bigg)
\nonumber\\
&\hspace{1.2cm} \times \langle g_s\bar{s}s\rangle ^2 \! + \! \frac{1}{\pi M^2}\bigg\{ \frac{1}{8} \! + \! \frac{m_{s}^{2}}{M^2}\,\bigg[ \frac{1}{3}+\frac{1}{2}\bigg(\ln \biggl(\frac{M^2}{\mu ^2} \biggr)\,+1\,-\gamma_E\bigg) \bigg] \bigg\}\, \langle \alpha _sG^2\rangle\,+\,\frac{m_{s}^{2}}{\pi ^2M^6}
\nonumber\\
&\hspace{1.2cm}\times\bigg[ -\frac{97}{162}\,+\frac{4}{27}\bigg(\ln \biggl(\frac{M^2}{\mu ^2} \biggr)+\frac{3}{2}-\gamma_E\bigg) \bigg] \langle g_{s}^{3}fG^3\rangle
\,+\,\frac{(2+\kappa ^2)}{M^4}\,\bigg\{ \frac{5}{54\pi ^2}\,+\,\frac{m_{s}^{2}}{\pi ^2M^2}
\nonumber\\
&\hspace{1.2cm}\times\bigg[ \frac{119}{729}-\frac{30}{253}\bigg(\ln \biggl(\frac{M^2}{\mu ^2} \biggr)+\frac{3}{2}-\gamma_E\bigg) \bigg] \bigg\}\langle g_{s}^{2}\bar{q}q\rangle ^2 +\frac{m_s}{M^4}\langle g_s\bar{s}\sigma TGs\rangle,
\label{Eq:twist-3s}
\\
&{\rm Im}I^{\rm pert}_{3;f_0}(s)=\frac{3s}{8\pi} \bigg\{\bigg(1 - \frac{m^2_s}{s}\bigg)\bigg[\frac{s^2 - (s - m^2_s)^2}{2s^2}+1\bigg]\bigg\}+\frac{3m^2_s}{4\pi}\bigg(\frac{2(1 - s)^2v^2}{s^2}\bigg),
\end{align}
\end{widetext}
where $v^2=1-\frac{4m^2_s}{s}$. As suggested in Ref.~\cite{Zhong:2021epq}, a better sum rules for $\langle\xi^{(s\bar s),n}_{2;f_0}\rangle$ should be
\begin{align}
\langle\xi^{(s\bar s),n}_{2;f_0}\rangle|_\mu=\frac{(\langle\xi^{p(s\bar s),0}_{3;f_0}\rangle \langle\xi^{(s\bar s),n}_{2;f_0}\rangle)|_{{\rm Form~ Eq.}\eqref{Eq:twist2s}}}{\sqrt{\langle\xi^{p(s\bar s),0}_{3;f_0}\rangle^2}|_{{\rm Form~Eq.}\eqref{Eq:twist-3s}}}
\label{Eq:xis}
\end{align}

Then we deal with the $\xi$-moments of $\phi^{(n\bar{n})}_{2;f_0}(x,\mu)$ for the second part. Its corresponding correlation function is
\begin{align}
\Pi^{(n\bar{n})}_{2;f_0}(z,q)&=i\int d^4xe^{iq\cdot x}\langle 0|T\{J^{(n\bar{n})}_n(x),\hat{J}_0^{(n\bar{n}),\dagger}(0)\}|0\rangle
\nonumber\\
&=(z\cdot q)^{n+1}I^{(n\bar{n})}_{2;f_0}(q^2)
\end{align}
with
\begin{align}
&J^{(n\bar{n})}_n(x)=\frac{1}{\sqrt 2}[{\bar u}(x) \DS z {(iz \cdot \tensor D )^n} u(x)+{\bar d}(x) \DS z {(iz \cdot \tensor D )^n} d(x)]
\nonumber\\
&\hat{J}_0^{(n\bar{n}),\dagger}(0)=\frac{1}{\sqrt 2}[{\bar u}(0)u(0)+{\bar d}(0)d(0)]
\end{align}
The detailed calculation process is consistent with the $s{\bar s}$ state, so there is no need to introduce it specifically here. The detailed calculation process can also refer to our previous work~\cite{Hu:2023pdl}. Since the current quark masses of $u$ and $d$ quarks are quite small, contributions from the $u$-and $d$-quark mass square terms can be safely neglected in the calculation. The final sum rules for the $\xi$-moments of $\phi^{(n\bar{n})}_{2;f_0}(x,\mu)$ becomes
\begin{widetext}
\begin{align}
&\frac{m_{f_0}{\bar f}^{2(n\bar{n})}_{f_0} \langle\xi^{p(n\bar{n}),0}_{3;f_0}\rangle \langle\xi^{(n\bar{n}),n}_{2;f_0}\rangle}{M^2 e^{m^2_{f_0}/M^2}}=\frac{[1-(-1)^n]}{2}\bigg\{\frac{1}{\pi}\frac{1}{M^2}\int^{s_{f_0}}_{(m_u+m_d)^2} dse^{-s/M^2}\bigg[-\frac{3(m_u+m_d)}{16\pi(n+1)(n+2)}
\nonumber\\
&\hspace{1.2cm}\times[2n+(-1)^n+3]\bigg]\,+ \frac{(\langle{\bar u}u\rangle+\langle{\bar d}d\rangle)}{M^2}-\frac{n(\langle g_s{\bar u}\sigma TGu\rangle+\langle g_s{\bar d}\sigma TGd\rangle)}{2 M^4}
+ \frac{2m_u\langle g_s{\bar u} u\rangle^2}{27M^6}
\nonumber\\
&\hspace{1.2cm}+ \frac{2m_d\langle g_s{\bar d} d\rangle^2}{27M^6}
+ L_{M^2}I_{\langle G^2\rangle} + L_{M^2}I_{\langle G^3\rangle} + L_{M^2}I_{\langle q^4\rangle}\bigg\}
\label{Eq:twist-2n}
\\
&L_{M^2} I_{\langle G^2\rangle}=-\frac{(m_u+m_d)\langle\alpha_sG^2\rangle}{48\pi M^4}\bigg\{  4\biggr[\ln \biggr( \frac{M^2}{\mu ^2}\!\biggl)\,+\,1-\gamma_E\biggl] \, +\, 2\,\,\frac{n + 2}{n + 1} \, -\, 4(\psi ^{(0)}(n + 1) + \gamma _E)
\nonumber\\
&\hspace{1.2cm} + \theta (n - 2) \biggl[ (6n + 4)\psi _3(n)  \,+ \frac{4n^3 + 10n^2 - 6n - 8}{n(n + 1)}\bigg] \,+ \theta (n - 1)\bigg[ - \frac{2}{n + 1} [ 3n^2+ 5n
\nonumber\\
&\hspace{1.2cm}  + 3+ (n + 1)\psi _2(n)]  - 8(n + 1)\bigg[\psi ^{(0)}(n + 1) + 2\gamma _E - \ln \bigg( \frac{M^2}{\mu ^2} \bigg)-1 \bigg]\bigg]\bigg\}
\\
&L_{M^2} I_{\rm \langle G^3\rangle}=\frac{(m_u+m_d)\langle g^3_sfG^3\rangle}{23040\pi^2 M^6}\bigg\{(1584 -7776n)\bigg[ \ln\bigg(\frac{M^2}{\mu^2} \bigg)+\frac{3}{2}-\gamma_E\bigg]  +\bigg[72(108n - 22)
\nonumber\\
&\hspace{1.2cm}
 -96(81n\,-\, 5)\bigg](\psi^{(0)}(n  + 1) + \gamma_E) -  \frac{12( - 1458n^2  -  1471n  -  121)} {n + 1}\,+  \theta(n-  2)\bigg[72
\nonumber\\
&\hspace{1.1cm}
\times(n - 1)\frac{ - 51n^3 - 35n^2  + 80n + 52} {n (n + 1)}  +   144(9n + 13) \psi_3(n) +  864n(n - 1)\bigg[\ln\bigg(\frac{M^2}{\mu^2} \bigg)
\nonumber\\
&\hspace{1.2cm}
+\frac{3}{2}-\gamma_E\bigg]\bigg]\,+ \theta(n  -  1)\bigg[288(5n + 8)\bigg[ \ln\bigg(\!\frac{M^2}{\mu^2}\! \bigg)+\frac{3}{2}-\gamma_E\bigg]\,-(1440n + 1824)(\psi^{(0)}
\nonumber\\
&\hspace{1.2cm}
 \times(n+ 1)+  \gamma_E)  -   \frac{12}{n+1}(431n^2 -1281n - 1030)-96\psi_2(n)\bigg]+ \theta(n - 3)\bigg[\frac{ 54}{n(n^2 - 1)}
\nonumber\\
&\hspace{1.2cm}
 \times(- 77n^5 - 249n^4+ 67n^3  -  95n^2  -  294n-  72)  -  216(20n^2 + 16n + 9)\psi_1(n)- 324
\nonumber\\
&\hspace{1.2cm}
\times(n^2+ 5n + 6)\bigg[\ln\bigg(\frac{M^2}{\mu^2}\bigg)+\frac{3}{2}-\gamma_E\bigg] + 648(n^2 + 5n +6)(\psi^{(0)}(n + 1) + \gamma_E)\bigg] \bigg\},
\\
&L_{M^2}I_{\langle q^4\rangle}=\frac{(m_u+m_d)(2+\kappa^2)\langle g^2_sq{\bar q}\rangle^2}{233280\pi^2 M^6}\,\,\bigg\{648(66n - 23)\,\bigg[\ln\bigg(\frac{M^2}{\mu^2}\bigg)\,+\,\frac{3}{2}\,-\,\gamma_E\bigg] \,+ \,14904
\nonumber\\
&\hspace{1.2cm}
 \times (\psi^{(0)}(n + 1)\, +\, \gamma_E)\,+ \,\frac{\,216(27n^2 \,+ 40n \,+ 82)}{n + 1}\,+\, \theta(n - 3)\bigg[\,3564(n^2 \,+\, 5n\,+\, 6)
\nonumber\\
&\hspace{1.2cm}
 \times\bigg[\ln\bigg(\frac{M^2}{\mu^2}\bigg)+\frac{3}{2}-\gamma_E\bigg] \,+ 1188(20n^2  +  16n  +  9)\psi_1(n) \,-    3564(n^2  +  5n+ 6)(\psi^{(0)}
\nonumber\\
&\hspace{1.2cm}
 \times(n+  1) +\gamma_E\,)\,+\,\frac{54 }{n(n^2 -  1)}(581n^5 + 2355n^4 + 1013n^3 -991n^2+606n+ 396)\bigg]
\nonumber\\
&\hspace{1.0cm}
   + \theta(n - 2)\,\bigg[ - 4752n(n - 1)\bigg[\ln\bigg(\frac{M^2}{\mu^2}\bigg)+\frac{3}{2}-\gamma_E\bigg] \,+  72(46n-  45)\psi_3(n) \,- 108
\nonumber\\
&\hspace{1.2cm}
\times(n - 1)\,\frac{\,- 82n^3 \,- 35n^2 \,+ 83n\, + 60}{n^2(n + 1)}\bigg] \,+\,\theta(n - 1)\bigg\{414\psi_2(n)\,- 648(3n \,- 10)
\nonumber\\
&\hspace{1.2cm}
 \times\bigg[(\psi^{(0)}(n + 1) + \gamma_E)\bigg]+ 648(3n - 10) \bigg[\ln\bigg(\frac{M^2}{\mu^2}\bigg)+\frac{3}{2}-\gamma_E\bigg]- \frac{18}{n + 1}(1555n^2
\nonumber\\
&\hspace{1.2cm}
  + 2283n + 368)\bigg\} \bigg\}.
\end{align}
The $0_{\rm th}$ moment expression of the twist-3 LCDA is
\begin{align}
&\frac{m^2_{f_0}{\bar f}^{2(n\bar{n})}_{f_0} \langle\xi^{p(n\bar{n}),0}_{3;f_0}\rangle^2}{M^2 e^{m^2_{f_0}/M^2}}=\frac{1}{2}\bigg\{\frac{1}{\pi}\frac{1}{M^2}\int^{s_{f_0}}_{(m_u+m_d)^2}\bigg(\frac{3s}{4\pi}\bigg) dse^{-s/M^2}-\frac{3(\langle{\bar u}u\rangle m_u+\langle{\bar d}d\rangle m_d)}{M^2}+\frac{\langle \alpha_sG^2\rangle}{4\pi M^2}
\nonumber\\
&\hspace{1.2cm}+\frac{5 (2+\kappa^2)\langle g_s^2 {\bar q}q\rangle^2}{27\pi^2M^4}+\frac{8(\langle g_s{\bar u}u\rangle^2+\langle g_s{\bar d}d\rangle^2)}{27M^4}+\frac{(\langle g_s{\bar u}\sigma TGu\rangle m_u+\langle g_s{\bar d}\sigma TGd\rangle m_d)}{M^4}\bigg\}
\label{Eq:twist-3n}
\end{align}
\end{widetext}
And similarly, the sum rules of $\langle\xi^{(n\bar{n}),n}_{2;f_0}\rangle$ is
\begin{align}
\langle\xi^{(n\bar{n}),n}_{2;f_0}\rangle|_\mu=\frac{(\langle\xi^{p(n\bar{n}),0}_{3;f_0}\rangle \langle\xi^{(n\bar{n}),n}_{2;f_0}\rangle)|_{{\rm Form~ Eq.}\eqref{Eq:twist-2n}}}{\sqrt{\langle\xi^{p(n\bar{n}),0}_{3;f_0}\rangle^2}|_{{\rm Form~Eq.}\eqref{Eq:twist-3n}}}
\label{Eq:xiq}
\end{align}
Therefore, the complete $\xi$-moment expression corresponding to the $f_0(980)$-meson twist-2 LCDA is
\begin{align}
\langle\xi^{n}_{2;f_0}\rangle|_\mu=\cos\theta\langle\xi^{(s\bar s),n}_{2;f_0}\rangle|_\mu+\sin\theta\langle\xi^{(n\bar{n}),n}_{2;f_0}\rangle|_\mu.
\label{Eq:xi}
\end{align}
The symmetry of the twist-3 LCDA for the $f_0(980)$-meson ensures the normalization of its zeroth moment, thereby enabling determination of $f_0(980)$-meson's decay constant by Eqs.~\eqref{Eq:twist-3s} and \eqref{Eq:twist-3n}. Analogous to the twist-2 LCDA analysis, this decay constant calculation requires separate calculation of two distinct components, {\it i.e.}
\begin{align}
{\bar f}_{f_0}=\cos\theta{\bar f}^{(s\bar s)}_{f_0}+\sin\theta{\bar f}^{(n\bar{n})}_{f_0}.
\end{align}

In Eqs.\eqref{Eq:twist2s}-\eqref{Eq:twist-3n}, the scale $\mu$ is the renormalization scale and $M^2$ is the Borel parameter. $m_s$, $m_u$, and $m_d$ are current quark masses of $s$-, $u$- and $d$-quark, respectively. $\langle {\bar s}s\rangle$, $\langle{\bar u}u\rangle$, $\langle{\bar d}d\rangle$ and $\langle g_ss{\bar s}\rangle^2$, $\langle g_su{\bar u}\rangle^2$, $\langle g_sd{\bar d}\rangle^2$ are double-quark condensates, $\langle g_s\bar{s}\sigma TGs\rangle$, $\langle g_s\bar{u}\sigma TGu\rangle$, $\langle g_s\bar{d}\sigma TGd\rangle$ are quark-gluon mixed condensates, $\langle g^2_ss{\bar s}\rangle^2$, $\langle g^2_s u{\bar u}\rangle^2$, $\langle g^2_s d{\bar d}\rangle^2$ are four-quark condensates, $\langle \alpha_sG^2\rangle$ is double-gluon condensate and $\langle g^3_s fG^3\rangle$ is triple-gulon condensate, respectively. The $0_{th}$-other derivative of the digamma function is given by $\psi^{(0)}(n+1)=\sum^n_{k=1}1/k-\gamma_E$ with the Euler constant $\gamma_E=0.577216$. As for the digamma function $\psi(n)$, we have
\begin{align}
\psi_1(n)&=\psi\bigg(\frac{n}{2}\bigg)-\psi\bigg(\frac{n-1}{2}\bigg)-(-1)^n\ln 4, \\
\psi_2(n)&=\psi\bigg(\frac{n-1}{2}\bigg)-\psi\bigg(\frac{n}{2}-1\bigg)+(-1)^n\ln 4, \\
\psi_3(n)&=\psi\bigg(\frac{n}{2}+1\bigg)-\psi\bigg(\frac{n+1}{2}\bigg)-(-1)^n\ln 4.
\end{align}
There is a linear relationship between $\langle\xi^{n}_{2;f_0}\rangle$ and $B_n$~\cite{Cheng:2023knr}, that is
\begin{align}
B_1(\mu)&=\frac{5}{3}\langle\xi^1_{2;f_0}\rangle|_\mu,  \\
B_3(\mu)&=\frac{3}{4}(7 \langle\xi^3_{2;f_0}\rangle|_\mu-3 \langle\xi^1_{2;f_0}\rangle|_\mu),  \\
&\quad\cdot\cdot\cdot
\label{Eq:bn}
\end{align}
Combining Eqs.~\eqref{Eq:da} and \eqref{Eq:xi}, the analytical expression of twits-2 LCDA can be obtained.

\section{Numerical Analysis}\label{sec:3}

\subsection{Behavior of $\xi$-moments, $f_{f_0}$ and $\phi_{2;f_0}$ for the $f_0(980)$-meson}

We are ready to derive the $\xi$-moments, the decay constant $f_{f_0}$ and the twist-2 LCDA $\phi_{2;f_0}$ of $f_0(980)$.

In calculation, we take the mass of $f_0(980)$ as $m_{f_0}=0.990\pm0.020~{\rm GeV}$, the $c$ current quark mass $m_c({\bar m}_c)=1.27\pm0.02{\rm GeV}$~\cite{PDG:2024}, the $u, d$ and $s$ current quark mass are adopted as $m_u=2.16^{+0.49}_{-0.26}~{\rm MeV}, m_d=4.67^{+0.48}_{-0.17}~{\rm MeV}$ and $m_s=93^{+11}_{-5}~{\rm MeV}$ at scale $\mu=2~{\rm GeV}$~\cite{PDG:2024}, respectively. The values of the scale dependent vacuum condensates are~\cite{Zhong:2021epq, Huang:2022xny, Wu:2022qqx}: $\langle {\bar u}u\rangle=\langle {\bar d}d\rangle=(-2.417^{+0.227}_{-0.114})\times10^{-2}{\rm GeV^3}$, $\langle {\bar s}s\rangle=\kappa\langle {\bar u}u\rangle$ with $\kappa=\langle {\bar s}s\rangle/\langle {\bar u}u\rangle=0.74\pm0.03$, $\langle g_s {\bar u}u\rangle^2=\langle g_s {\bar d}d\rangle^2=(2.082^{+0.734}_{-0.679})\times10^{-3}{\rm GeV^6}$, $\langle g_s{\bar s}s\rangle^2=\kappa^2\langle g_s {\bar u}u\rangle^2$, $\langle g_s{\bar u}\sigma TGu\rangle=(-1.934^{+0.188}_{-0.103})\times10^{-2}{\rm GeV^5}$, $\langle g_ss\sigma TG{\bar s}\rangle=\kappa\langle g_s{\bar u}\sigma TGu\rangle$, $\langle g^2_sq{\bar q}\rangle^2=(7.420^{+2.614}_{-2.483})\times 10^{-3}{\rm GeV^6}$ at $\mu=2~{\rm GeV}$. the values of the scale independent gluon condensates are: $\langle \alpha_sG^2\rangle=0.038\pm0.011~{\rm GeV^4}$ and $\langle g^3_sfG^3\rangle=0.045~{\rm GeV^6}$. When doing the numerical calculation, each vacuum condensates and current quark masses should be run from their initial values at an initial scale to the required scale by applying the renormalization group equations (RGEs)~\cite{Zhong:2021epq, Yang:1993bp}.

\begin{figure}[htb!]
\begin{center}
\includegraphics[width=0.45\textwidth]{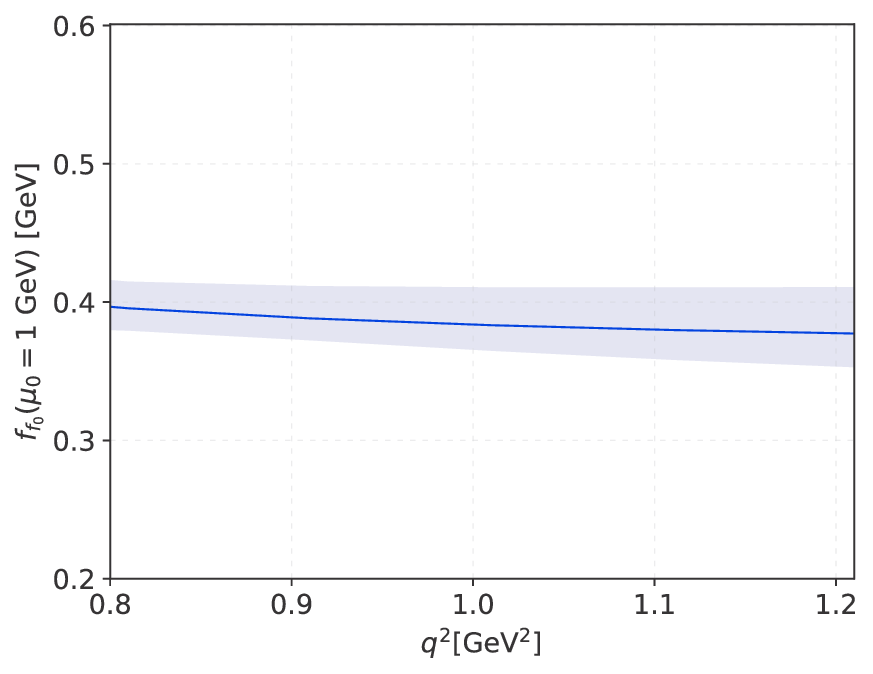}
\end{center}
\caption{The curve of the $f_0(980)$ decay constant $f_{f_0}$ within the Borel parameter $M^2$ with $\mu_0=1~{\rm GeV}$, where the shaded band indicated the uncertainties form the input parameters.} \label{fig:ff0}
\end{figure}
The continuum threshold parameter $s_{f_0}$ and Borel parameter $M^2$ are two important parameters for determining the decay constant $f_{f_0}$. Usually, the continuum threshold $s_{f_0}$ is taken as the value that is close to the squared mass of the first excited state of $f_0(980)$. As a conservative prediction, we set $s_{f_0}=1.9\pm0.1~{\rm GeV^2}$ to do our discussion, whose central value is set by $f_0(1370)$~\cite{PDG:2024}. In addition, it is also necessary to require that $\langle\xi^{p,0}_{3;f_0}\rangle$ be normalized. To determine the allowable $M^2$ range, we require that $f_{f_0}$ is flat within the allowable Borel window. The $f_0(980)$ decay constant $f_{f_0}$ versus the Borel parameter $M^2$ is shown in Fig.~\ref{fig:ff0}. Besides, the contribution of the continuum state does not exceed $30\%$, and the six-dimensional condensation does not exceed $1\%$. Therefore, we obtain the range of the Borel parameter as $M^2\in[0.816, 1.150]{\rm GeV^2}$. Since $f_{f_0}$ exhibits scale dependence~\cite{Cheng:2005nb}, we present here only the results evaluated at the initial energy scale $(\mu_0=1{\rm GeV})$.

\begin{table}[htb!]
\renewcommand\arraystretch{1.3}
\center
\small
\caption{The decay constant $f_{f_0}$ (in unit: ${\rm GeV}$) using the BFTSR approach. As a comparison, typical result derived from other approaches have also been presented. } \label{Tab:ff0}
\begin{tabular}{l c }
\hline
~~~~~~~~~~~~~~~~~~~~~~~~~  & ~~~$f_{f_0}(\mu_0={\rm 1 GeV})$~~~  \\      \hline
This work                 & $0.386\pm0.009$        \\
QCD SR'06~\cite{Cheng:2005nb}          & $0.370\pm0.020$       \\
3PSR~\cite{DeFazio:2001uc}               & $0.180\pm0.015$       \\ \hline
\end{tabular}
\end{table}

For comparison, theoretical predictions from two other approaches are also listed in Table \ref{Tab:ff0}. Ref.~\cite{Cheng:2005nb} employs a method based on ground state and excited state separation for its calculations. Furthermore, the associated correlation function they constructed is different from ours. Notably, the $f_0(980)$-meson has been treated as a pure $s\bar{s}$ state in their frameworks. It can be found that our prediction result is close to that of QCD SR'06~\cite{Cheng:2005nb} and there is a gap compared with the prediction results of 3PSR~\cite{DeFazio:2001uc}. This further verifies the predicted results of the experiment, e.g. the $s\bar{s}$ component dominates, while the $n\bar{n}$ component cannot be ignored.

After determining the decay constant, we then calculate the moments of the twist-2 LCDA $\phi_{2;f_0}$. Given the antisymmetric nature of the twist-2 LCDA, its continuum threshold must be calibrated using the zeroth moment of the twist-3 LCDA, maintaining consistency with the continuum threshold employed in the calculation of decay constant.
\begin{figure}[htb]
\begin{center}
\includegraphics[width=0.45\textwidth]{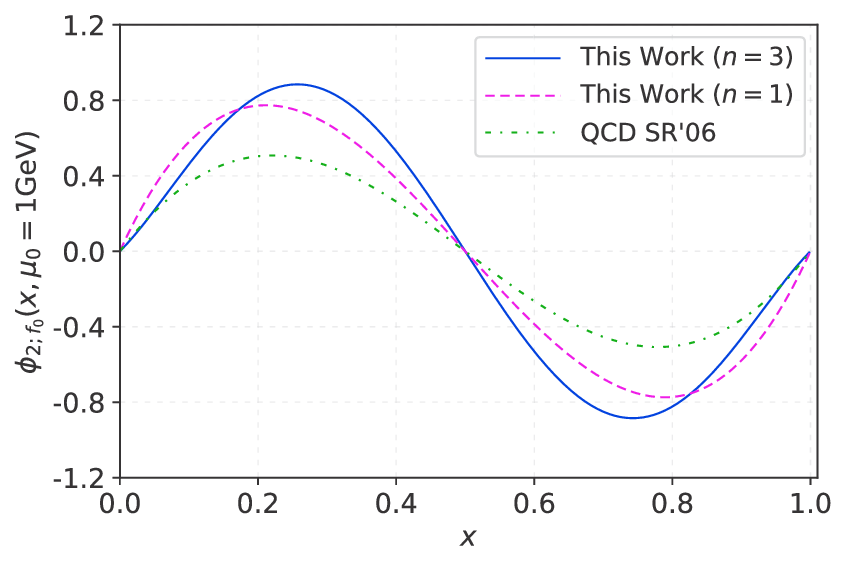}
\end{center}
\caption{The $f_0(980)$ twist-2 LCDA $\phi_{2;f_0}(x,\mu_0=1{\rm GeV})$. As a comparsion, the QCD SR'06~\cite{Cheng:2005nb} predicted curve is also given.} \label{fig:phi2}
\end{figure}
Through analysis of contributions from continuum states and six-dimensional condensates, constraints on the $M^2$ parameter are established: for the first two moments of the twist-2 LCDA, the continuum contributes are less than $30\%$ and $35\%$, respectively, while the six-dimensional condensate contributes are less than $1\%$ and $1\%$, respectively. This leads to the ranges of $M^2$ as $[1.046, 1.689]~{\rm GeV^2}$ (for the first-order moment) and $[1.204, 2.141]{\rm GeV^2}$ (for the second-order moment). By taking all uncertainty sources into consideration and applying the RGEs of the moments, $\langle\xi^{n}_{2;f_0}\rangle|_\mu$ at the initial scale $\mu_0$ and the special scale $\mu_k=\sqrt{m^2_{D^+_s}-m^2_c}\approx 1.5{\rm GeV}$ are
\begin{align}
\langle\xi^1_{2;f_0}\rangle|_{\mu_0} &=-0.693^{+0.095}_{-0.081}, \\
\langle\xi^1_{2;f_0}\rangle|_{\mu_k} &=-0.542^{+0.074}_{-0.063}, \\
\langle\xi^3_{2;f_0}\rangle|_{\mu_0} &=-0.261^{+0.097}_{-0.095}, \\
\langle\xi^3_{2;f_0}\rangle|_{\mu_k} &=-0.217^{+0.079}_{-0.063}.
\end{align}
And the corresponding Gegenbauer moments are
\begin{align}
B_1(\mu_0) &=-1.156^{+0.157}_{-0.133},  \\
B_1(\mu_k) &=-0.903^{+0.123}_{-0.104},  \\
B_3(\mu_0) &=0.191^{+0.294}_{-0.317},  \\
B_3(\mu_k) &=0.079^{+0.252}_{-0.269}.
\end{align}
Substituting the determined parameters into Eq.~\eqref{Eq:da}, the analytical form of the twist-2 LCDA can be obtained. To facilitate visual comparison, Fig.~\ref{fig:phi2} displays the curves of twist-2 LCDA $\phi_{2;f_0}(x,\mu_0)$ for both $n=1$ and $n=3$ cases, along with the theoretical predictions obtained from QCD SR'06~\cite{Cheng:2005nb}. The first two Gegenbauer moments predicted by QCD SR'06~\cite{Cheng:2005nb} are $b_1=-0.78\pm0.08$ and $b_3=0.02\pm0.07$ respectively. Our predicted behavior of the twist-2 LCDA exhibits good consistency with the theoretical results from QCD SR'06, both of which exhibiting antisymmetric characteristics.

\subsection{The TFF and branching fraction for the semi-leptonic $D^+_s\to f_0(980)(\to \pi^+\pi^-)e^+\nu_e$}

In order to calculate the TFFs of the $D^+_s\to f_0(980)$ decay in Eq.~\eqref{Eq:fp}, we adopt the heavy meson mass $m_{D^+_s}=1.9685{\rm GeV}$~\cite{PDG:2024}, the decay constants $f_{D^+_s}=0.258\pm0.005{\rm GeV}$~\cite{Duplancic:2015zna}, and $f_{f_0}(\mu_k)=0.440\pm0.010$ that is obtained by using the BFTSR. The parameter values needed when calculating the BF for the semi-leptonic $D^+_s\to f_0(980)(\to \pi^+\pi^-)e^+\nu_e$ are: $m_\pi\approx 0.140~{\rm GeV}$, $m_K\approx0.494~{\rm GeV}$, $\tau_{D_s}=0.5012~{\rm ps}$, $m_e=0.511\times10^{-3}~{\rm GeV}$, $G_F=1.166\times10^{-5}~{\rm GeV^{-2}}$. The above parameters are all from the PDG~\cite{PDG:2024}. Besides, it is also necessary to know $g_1=165~{\rm MeV}/c^2$ and $g_2=4.21 g_1$~\cite{BES:2004twe, BESIII:2016tqo}. For the effective threshold parameter and Borel parameter, we take $s_0=6.5\pm0.5~{\rm GeV}^2$ and $M^2=7\pm1~{\rm GeV^2}$.

\begin{table}[htb]
\renewcommand\arraystretch{1.3}
\center
\small
\caption{Theoretical predictions on the TFF $f_+(0)$ at the large recoil point $q^2=0$. } \label{Tab:fp}
\begin{tabular}{l c}
\hline
~~~~~~~~~~~~~~~~~~~~~~~~~~~~  & ~~~$f_+ (0)$~~~~   \\
\hline
This work (LCSR)   & $0.516_{-0.024}^{+0.027}$     \\
BESIII'24~\cite{BESIII:2023wgr}          & $0.518\pm0.018\pm0.036$     \\
LCSR'10~\cite{Colangelo:2010bg}         & $0.300\pm0.030$      \\
3PSR'04~\cite{Bediaga:2003zh}             & $0.50\pm0.13$      \\
3PSR'10~\cite{Aliev:2007uu}            & $0.48\pm0.23$                \\
LCSR'24-s1~\cite{Cheng:2023knr}       & $0.580\pm 0.070$     \\
LCSR'24-s2~\cite{Cheng:2023knr}       & $0.400\pm 0.060$      \\
LCSR'24-s3~\cite{Cheng:2023knr}      & $0.780_{-0.100}^{+0.130}$         \\
\hline
\end{tabular}
\end{table}

With the above input and the $f_0(980)$ twist-2 LCDA determined in the previous subsection, the dependence of the TFFs for the semi-leptonic $D^+_s \to f_0(980)$ decay on $q^2$ in the valid $q^2$ regions of the LCSR can be obtained. In particular, the values of these TFFs at the large recoil point are exhibited in Table~\ref{Tab:fp}. As a comparison, we have also presented the results derived from various theoretical approaches and the recent BESIII result in Table~\ref{Tab:fp}. Owing to substantial discrepancies among the predictions from various theoretical frameworks and experimental results  e.g., the LFQM calculation yielding $f_+ (0) = 0.24 \pm 0.05$~\cite{Ke:2009ed}, which deviates from experimental predictions by approximately $46\%$ we primarily present the theoretical results derived from the LCSR and 3PSR methods in this table. It can be observed that our results are in close agreement with those of the BESIII Collaboration~\cite{BESIII:2023wgr}. A recent LCSR study~\cite{Cheng:2023knr} shows that the predictions vary with different choices of current (s1, s2, s3). Compared with the s2 and s3 models, the prediction for s1 is closer to the BESIII result. This behavior is attributed to the fact that s1 uses the traditional current, which is consistent with the associated current we constructed.

\begin{table}[htb]
\renewcommand\arraystretch{1.3}
\center
\small
\caption{Fitting parameters $\alpha_1$ and $\alpha_2$ for the central TFF $f_+^{(C)}(q^2)$, the upper $f_+^{(U)}(q^2)$ and the lower $f_+^{(L)}(q^2)$. $\Delta$ is the measure of
the quality of extrapolation. } \label{Tab:fit}
\begin{tabular}{l c c c c}
\hline
~~~~~~~~~& ~~~$f_+^{(C)}(q^2)$~~~ & ~~~$f_+^{(U)}(q^2)$~~~ & ~~~$f_+^{(L)}(q^2)$~~~   \\
\hline
$\alpha_1$      & $-1.139$    & $-1.646$    & $-0.761$    \\
$\alpha_2$      & $-65.777$   & $-62.603$   & $-72.413$    \\
$\Delta$       & $0.090\%$    & $0.091\%$   & $0.097\%$      \\
\hline
\end{tabular}
\end{table}

Theoretically, the LCSR approach for $D^+_s\to f_0(980)$ TFFs are relable in low and intermediate $q^2$-regions, which can be extrapolated to all the physically allowable region $m^2_\ell\ll(m_{D^+_s}-m_{f_0})^2\approx0.96 {\rm GeV^2}$. In the context of analytic continuation, various technical approaches are available, including but not limited to series expansion (SE), simplified series expansion (SSE), single-pole (SP), and double-pole parametrization, etc. In the present paper, we adopt the approach of SSE to do the extrapolation. one of the advantages of this is the simplicity to translate the near-threshold behavior of the form factors into a useful constraint on the expansion coefficients. So the TFFs take the following form~\cite{Bharucha:2010im}
\begin{align}
f_+(q^2)=\frac{1}{1-q^2/m^2_{D^*_s}}\sum^2_{k=1}\alpha_kz^k(t, t_0)
\end{align}
The function $z(t=q^2, t_0)$, which incorporates the parameters $t_\pm, t_0$ ang $t$, is defined as
\begin{align}
z(t, t_0)=\frac{\sqrt{t_+-t}-\sqrt{t_+-t_0}}{\sqrt{t_+-t}+\sqrt{t_+-t_0}}
\end{align}
where $t_\pm=(m_{D^+_s}\pm m_{f_0})^2$ and $t_0=t_+(1-\sqrt{1-t_-/t_+})$. In this approach, the simple pole $1-q^2/m^2_{D^*_s}$ is used to account for the low-lying resonances, and $m_{D^*_{s}}=2.3177~{\rm GeV}$ is $D_s^+$ meson resonances~\cite{PDG:2024, Cheng:2023knr}. The free parameters $\alpha_1$ and $\alpha_2$ are fixed to make the $\Delta$ as small as possible, such as $\Delta<1\%$, where $\Delta$ is used to measure the quality of extrapolation and is define as
\begin{align}
\Delta=\frac{\sum_t|F_+(t)-F^{\rm fit}_+(t)|}{\sum_t|F_+(t)|}\times 100
\end{align}
where $t\in[0, \frac{1}{40},\cdot\cdot\cdot, \frac{40}{40}]\times 0.58~{\rm GeV}$. The fitting parameters $\alpha_1, \alpha_2$ for the TFF $f_+(q^2)$ and the quality-of-fit $\Delta$ in Table~\ref{Tab:fit}. It shows that under those choices of parameters, all the $\Delta$ values are
no more than $0.097\%$, indicating a good agreement of the extrapolated curves with the LCSR within the same $q^2$-region of $q^2\in[0, 0.58]~{\rm GeV^2}$.

\begin{figure}[htb]
\begin{center}
\includegraphics[width=0.45\textwidth]{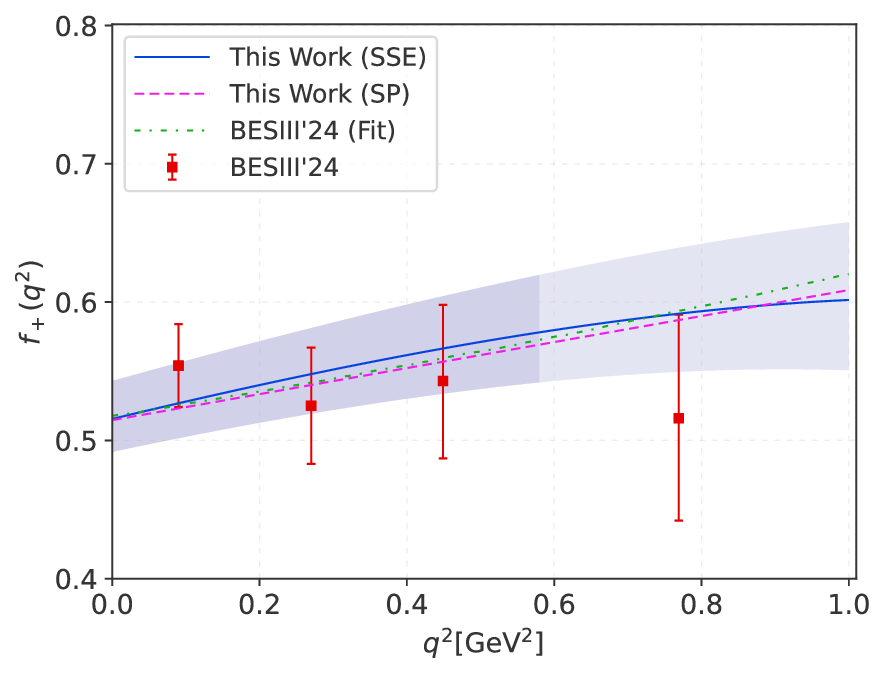}
\end{center}
\caption{The extrapolated TFF $f_+(q^2)$ in entire $q^2$-region, where the solid line is the central value and the shaded band shows its uncertainty. The thicker shaded band shows the LCSR prediction. As a comparison, BESIII'24 and BESIII'24 fit are also presented.} \label{fig:fp}
\end{figure}

We present the TFF $f_+(q^2)$ across the entire momentum transfer $q^2$ range in Fig.~\ref{fig:fp}, along with the latest experimental measurements for direct comparison. To facilitate a more comprehensive comparison with experimental results, we also include the results of analytic continuation using the SP parametrization~\cite{Achasov:2020qfx, Becirevic:1999kt}, with the central value of $f_+(q^2)$ plotted in Fig.~\ref{fig:fp}. While the SP parametrization yields predictions closer to experimental data than the SSE parametrization, it relies solely on endpoint values of the TFF and neglects information from the medium- and low-energy regions accessible within the LCSR framework. Consequently, we recommend the SSE parametrization for extrapolation. Across the entire $q^2$ range, our theoretical predictions represented by their central values are in agreement with the BESIII'24 fit data and lie within the uncertainty bands of the latest BESIII'24 results.

\begin{figure}[htb]
\begin{center}
\includegraphics[width=0.45\textwidth]{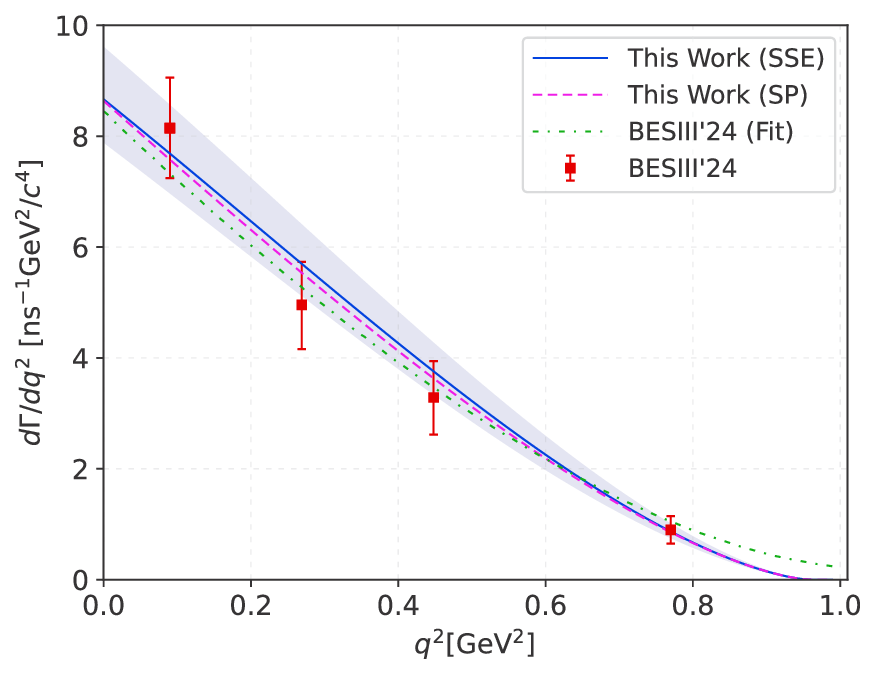}
\end{center}
\caption{Differential decay width of the semi-leptonic $D^+_s\to f_0(980)(\to \pi^+\pi^-)e^+\nu_e$ decay. As a comparison, BESIII'24 and BESIII'24 fit are also presented.} \label{fig:dgamma}
\end{figure}

Then, we can calculate the differential decay width of the semi-leptonic decay $D^+_s\to f_0(980)(\to \pi^+\pi^-)e^+\nu_e$ by using Eq.~\eqref{Eq:dgamma}. In calculation, we take CKM matrix element $|V_{cs}|=0.974$~\cite{BESIII:2023wgr}. The resulted differential decay width for the semi-leptonic $D^+_s\to f_0(980)(\to \pi^+\pi^-)e^+\nu_e$ decay versus $q^2$ is shown in Fig.~\ref{fig:dgamma}, where the solid line is for the central value and the shaded band shows the corresponding error. As a comparison, BESIII'24 and BESIII'24 fit are also presented. Fig.~\ref{fig:dgamma} shows that the central values and errors we calculated are all within the predicted error range of the latest BESIII'24 experiment. In different to BESIII'24 fit, our prediction tends to $0$ when $q^2\to (m_{D_s}-m_{f_0})^2\approx0.96 {\rm GeV^2}$, which is similar to most of the other semi-leptonic decay processes such as final state involving $a_0(980)$, $K^*_0(1430)$, $\pi$, $K$, and etc. The primary reason for the discrepancy between our predicted differential decay width and the BESIII'24 fit results lies in the selection of $\pi^+\pi^-$ invariant mass ($s$) ranges. During the BESIII'24 fitting process, the variable $s\in[0.6, 1.6]~{\rm GeV}/c^2$~\cite{BESIII:2023wgr}, whereas our actual computational framework restricts this interval to $s\in[0.9, 0.96]~{\rm GeV}/c^2$~\cite{Wang:2016wpc, BES:2004twe}.

\begin{table}[htb]
\renewcommand\arraystretch{1.3}
\center
\small
\caption{The branching fraction of $D^+_s\to f_0(980)(\to \pi^+\pi^-)e^+\nu_e$, where the BESIII and CLEO results are given as a comparison. } \label{Tab:BF}
\begin{tabular}{l c }
\hline
~~~~~~~~~~~~~~~~~  & ~~~${\mathcal B}(D_s\to f_0(980)(\to \pi^+\pi^-)e^+\nu_e)$~~~  \\      \hline
This work(SSE)          & $(1.783^{+0.227}_{-0.189})\times 10^{-3}$      \\
This work(SP)          & $(1.743^{+0.190}_{-0.157})\times 10^{-3}$      \\
CLEO'09~\cite{CLEO:2009dyb}          & $(1.3\pm0.4\pm0.1)\times 10^{-3}$    \\
CLEO'09~\cite{CLEO:2009ugx}          & $(2.0\pm0.3\pm0.1)\times 10^{-3}$    \\
BESIII'24~\cite{BESIII:2023wgr}          & $(1.72\pm0.13\pm0.10)\times 10^{-3}$      \\
\hline
\end{tabular}
\end{table}

Finally, by using the lifetimes of $D^+_s$-meson, we calculate the branching fraction of $D^+_s\to f_0(980)(\to \pi^+\pi^-)e^+\nu_e$ by using the extrapolated TFFs under the SSE and SP methods, respectively, which are listed in Table~\ref{Tab:BF}. Meanwhile, the BESIII and CLEO experimental predictions~\cite{BESIII:2023wgr, CLEO:2009ugx, CLEO:2009dyb} are also given as a comparison. Our results are consistent with the latest experimental results within errors, with only a $3.7\%$ shift of the central value for the SSE extrapolation, and a $1.3\%$ shift of the central value for the SP extrapolation, accordingly.

\section{Summary}\label{sec:summary}

In this work, we treat the $f_0(980)$ as a mixed state composed of distinct quark components, $|s{\bar s}\rangle\cos\theta+|n{\bar n}\rangle\sin\theta$, and employ the BFTSR method to compute its decay constant and twist-2 LCDA. The predicted decay constant, $f_{f_0}|_{1\;{\rm GeV}} = 0.386\pm0.009 \, \text{GeV}$, aligns closely to the QCD SR'06 result, which also validates the latest BESIII data. For the $f_0(980)$-meson, its $s\bar{s}$ component dominates, while the $n\bar{n}$ component remains substantial and non-negligible. In Fig.~\ref{fig:phi2}, we present the twist-2 LCDAs for $n=1$ and $n=3$ at the initial scale of $1$ GeV, both of which exhibit antisymmetry.

Subsequently, we calculate the TFF $f_+(0)$ for $D^+_s\to f_0(980)$, observing that the endpoint results align precisely with the latest BESIII experimental predictions. By further extending the TFF analysis to the full momentum transfer $q^2$ range through SSE parametrization, we derive the differential decay width. The final predictions are consistent with the BESIII's latest theoretical prediction across the entire $q^2$ region. Since experimental results are presented using the SP parameterization, we also include the SP predictions here. As observed, the SP parameterization predicted TFF and BF values closer to experimental measurements. However, due to the inherent limitations of the SP approach, we recommend prioritizing the SSE parameterization results for theoretical analysis.

Finally, based on the width formula given in the latest BESIII experiment, which includes the Flatt\'{e} factor, we calculated the differential decay width of $D^+_s\to f_0(980)(\to \pi^+\pi^-)e^+\nu_e$ (Fig.~\ref{fig:dgamma}) and the branching fraction (Table~\ref{Tab:BF}). It is found that there is a gap between our final prediction result and the experiment, mainly because the $s$-range is different. However, our predicted value is within the error range of the experiment. Among them, the central value of the branching fraction predicted by the SSE parameterization has an error of only $3.7\%$ compared with the BESIII 24's predictions. It is hoped that these results can provide important insights into the intrinsic properties of the light scalar $f_0(980)$-meson and the non-perturbation mechanism controlling the decay dynamics of the charm meson.

\section{Acknowledgments}

Tao Zhong and Hai-Bing Fu would like to thank the Institute of High Energy Physics of Chinese Academy of Sciences for their warm and kind hospitality. This work was supported in part by the National Natural Science Foundation of China under Grant No.12347101, No.12175025, No.12265009 and No.12265010, the Project of Guizhou Provincial Department of Science and Technology under Grant No.ZK[2023]024, No.MS[2025]219, No.CXTD[2025]030, the Graduate Research and Innovation Foundation of Chongqing, China under Grant No.ydstd1912, the Fundamental Research Funds for the Central Universities under Grant No.2024IAIS-ZD009.


\begin{thebibliography}{99}

\bibitem{Black:1998wt}
  D.~Black, A.~H.~Fariborz, F.~Sannino and J.~Schechter,
  Putative light scalar nonet,
  \href{https://doi.org/10.1103/PhysRevD.59.074026}
  {Phys. Rev. D \textbf{59} (1999), 074026}.

\bibitem{Achasov:2012kk}
  N.~N.~Achasov and A.~V.~Kiselev,
  Light scalars in semi-leptonic decays of heavy quarkonia,
  \href{https://doi.org/10.1103/PhysRevD.86.114010}
  {Phys. Rev. D \textbf{86} (2012), 114010}.

\bibitem{Sekihara:2015iha}
  T.~Sekihara and E.~Oset,
  Investigating the nature of light scalar mesons with semileptonic decays of $D$ mesons,
  \href{https://doi.org/10.1103/PhysRevD.92.054038}
  {Phys. Rev. D \textbf{92} (2015), 054038}.

\bibitem{Soni:2020sgn}
  N.~R.~Soni, A.~N.~Gadaria, J.~J.~Patel and J.~N.~Pandya,
  Semileptonic Decays of Charmed Mesons to Light Scalar Mesons,
  \href{https://doi.org/10.1103/PhysRevD.102.016013}
  {Phys. Rev. D \textbf{102} (2020), 016013}.

\bibitem{PDG:2024}
  S. Navas et al. (Particle Data Group),
  Review of Particle Physics,
  \href{https://doi.org/10.1103/PhysRevD.110.030001}
  {Phys. Rev. D \textbf{110}, 030001 (2024)}.

\bibitem{Tornqvist:1995kr}
  N.~A.~Tornqvist,
  Understanding the scalar meson $q\bar{q}$ nonet,
  \href{https://doi.org/10.1007/BF01565264}
  {Z. Phys. C \textbf{68} (1995), 647-660}.

\bibitem{vanBeveren:1998qe}
  E.~van Beveren and G.~Rupp,
  Comment on `Understanding the scalar meson $q\bar{q}$ nonet',
  \href{https://doi.org/10.1007/s100520050769}
  {Eur. Phys. J. C \textbf{10} (1999), 469-472}.

\bibitem{Morgan:1993td}
  D.~Morgan and M.~R.~Pennington,
  New data on the $K{\bar K}$ threshold region and the nature of the $f_0(S^*)$,
  \href{https://doi.org/10.1103/PhysRevD.48.1185}
  {Phys. Rev. D \textbf{48} (1993), 1185-1204}.

\bibitem{Tornqvist:1995ay}
  N.~A.~Tornqvist and M.~Roos,
  Resurrection of the sigma meson,
  \href{https://doi.org/10.1103/PhysRevLett.76.1575}
  {Phys. Rev. Lett. \textbf{76} (1996), 1575-1578}.

\bibitem{Ke:2009ed}
  H.~W.~Ke, X.~Q.~Li and Z.~T.~Wei,
  Whether new data on $D_s\to f_0(980)e^+\nu_e$ can be understood if $f_0(980)$ consists of only the conventional $q{\bar q}$ structure,
  \href{https://doi.org/10.1103/PhysRevD.80.074030}
  {Phys. Rev. D \textbf{80} (2009), 074030}.

\bibitem{Jaffe:1976ig}
  R.~L.~Jaffe,
  Multi-Quark Hadrons. 1. The Phenomenology of $Q^2{\bar Q}^2$ Mesons,
  \href{https://doi.org/10.1103/PhysRevD.15.267}
  {Phys. Rev. D \textbf{15} (1977), 267}.

\bibitem{Alford:2000mm}
  M.~G.~Alford and R.~L.~Jaffe,
  Insight into the scalar mesons from a lattice calculation,
  \href{https://doi.org/10.1016/S0550-3213(00)00155-3}
  {Nucl. Phys. B \textbf{578} (2000), 367-382}.

\bibitem{Fariborz:2009cq}
  A.~H.~Fariborz, R.~Jora and J.~Schechter,
  Global aspects of the scalar meson puzzle,
  \href{https://doi.org/10.1103/PhysRevD.79.074014}
  {Phys. Rev. D \textbf{79} (2009), 074014}.

\bibitem{Brito:2004tv}
  T.~V.~Brito, F.~S.~Navarra, M.~Nielsen and M.~E.~Bracco,
  QCD sum rule approach for the light scalar mesons as four-quark states,
  \href{https://doi.org/10.1016/j.physletb.2005.01.008}
  {Phys. Lett. B \textbf{608} (2005), 69-76}.

\bibitem{Achasov:2010fh}
  N.~N.~Achasov and A.~V.~Kiselev,
  The analytical $\pi\pi$ scattering amplitude and the light scalars,
  \href{https://doi.org/10.1103/PhysRevD.83.054008}
  {Phys. Rev. D \textbf{83} (2011), 054008}.

\bibitem{Kim:2017yvd}
  H.~Kim, K.~S.~Kim, M.~K.~Cheoun and M.~Oka,
  Tetraquark mixing framework for isoscalar resonances in light mesons,
  \href{https://doi.org/10.1103/PhysRevD.97.094005}
  {Phys. Rev. D \textbf{97} (2018), 094005}.

\bibitem{Baru:2003qq}
   V.~Baru, J.~Haidenbauer, C.~Hanhart, Y.~Kalashnikova and A.~E.~Kudryavtsev,
  Evidence that the $a_0(980)$ and $f_0(980)$ are not elementary particles,
  \href{https://doi.org/10.1016/j.physletb.2004.01.088}
  {Phys. Lett. B \textbf{586} (2004), 53-61}.

\bibitem{Branz:2007xp}
  T.~Branz, T.~Gutsche and V.~E.~Lyubovitskij,
  $f_0(980)$-meson as a $K\bar{K}$ molecule in a phenomenological Lagrangian approach,
  \href{https://doi.org/10.1140/epja/i2008-10635-1}
   {Eur. Phys. J. A \textbf{37} (2008), 303-317}.

\bibitem{Lee:2013mfa}
  H.~J.~Lee, N.~I.~Kochelev and Y.~Oh,
  QCD sum rule study on the $f_0$(980) structure as a pure $K\bar{K}$ bound state,
  \href{https://doi.org/10.1103/PhysRevD.87.117901}
  {Phys. Rev. D \textbf{87} (2013), 117901}.

\bibitem{Robson:1977pm}
  D.~Robson,
  A Basic Guide for the Glueball Spotter,
  \href{https://doi.org/10.1016/0550-3213(77)90110-9}
  {Nucl. Phys. B \textbf{130} (1977), 328-348}.

\bibitem{Narison:1996fm}
  S.~Narison,
  Masses, decays and mixings of gluonia in QCD,
  \href{https://doi.org/10.1016/S0550-3213(97)00562-2}
  {Nucl. Phys. B \textbf{509} (1998), 312-356}.

\bibitem{Lebiedowicz:2020bwo}
  P.~Lebiedowicz, R.~Maciu\l{}a and A.~Szczurek,
  Production of $f_{0}(980)$ meson at the LHC: Color evaporation versus color-singlet gluon-gluon fusion,
  \href{https://doi.org/10.1016/j.physletb.2020.135475}
  {Phys. Lett. B \textbf{806} (2020), 135475}.

\bibitem{BaBar:2008gpr}
  B.~Aubert \textit{et al.} [BaBar Collaboration],
  Study of the decay $D^+_{s} \to K^{+} K^{-} e^{+} \nu_{e}$,
  \href{https://doi.org/10.1103/PhysRevD.78.051101}
  {Phys. Rev. D \textbf{78} (2008), 051101}.

\bibitem{CLEO:2009dyb}
  J.~Yelton \textit{et al.} [CLEO Collaboration],
  Absolute Branching Fraction Measurements for Exclusive $D_{(s)}$ Semileptonic Decays,
  \href{https://doi.org/10.1103/PhysRevD.80.052007}
  {Phys. Rev. D \textbf{80} (2009), 052007}.

\bibitem{CLEO:2009ugx}
  K.~M.~Ecklund \textit{et al.} [CLEO Collaboration],
  Study of the semileptonic decay $D^+_s\to f_0(980)e^+\nu_e$ and implications for $B^0_s\to J/\psi f_0$,
  \href{https://doi.org/10.1103/PhysRevD.80.052009}
  {Phys. Rev. D \textbf{80} (2009), 052009}.

\bibitem{BESIII:2018qmf}
  M.~Ablikim \textit{et al.} [BESIII Collaboration],
  Observation of $D^+ \to f_0(500) e^+\nu_e$ and Improved Measurements of $D \to\rho e^+\nu_e$,
  \href{https://doi.org/10.1103/PhysRevLett.122.062001}
  {Phys. Rev. Lett. \textbf{122} (2019), 062001}.

\bibitem{BESIII:2021drk}
  M.~Ablikim \textit{et al.} [BESIII Collaboration],
  Study of light scalar mesons through $D^+_s \to \pi^0\pi^0e^+ \nu_e$ and $K^0_S K^0_S e^+ \nu_e$ decays,
  \href{https://doi.org/10.1103/PhysRevD.105.L031101}
  {Phys. Rev. D \textbf{105} (2022), L031101}.

\bibitem{BESIII:2023wgr}
  M.~Ablikim \textit{et al.} [BESIII Collaboration],
  Study of the $f_0(980)$ and $f_0(500)$ Scalar Mesons through the Decay $D_s^+\to\pi^+\pi^-e^+\nu_e$,
  \href{https://doi.org/10.1103/PhysRevLett.132.141901}
  {Phys. Rev. Lett. \textbf{132} (2024), 141901}.

\bibitem{Colangelo:2010bg}
  P.~Colangelo, F.~De Fazio and W.~Wang,
  $B_s\to f_0(980)$ form factors and $B_s$ decays into $f_0(980)$,
  \href{https://doi.org/10.1103/PhysRevD.81.074001}
  {Phys. Rev. D \textbf{81} (2010), 074001}.

\bibitem{Shi:2015kha}
  Y.~J.~Shi and W.~Wang,
  Chiral Dynamics and S-wave contributions in Semileptonic $D_s/B_s$ decays into $\pi^+\pi^-$,
  \href{https://doi.org/10.1103/PhysRevD.92.074038}
  {Phys. Rev. D \textbf{92} (2015), 074038}.


\bibitem{El-Bennich:2008rkp}
  B.~El-Bennich, O.~Leitner, J.~P.~Dedonder and B.~Loiseau,
  The Scalar Meson $f_0(980)$ in Heavy-Meson Decays,
  \href{https://doi.org/10.1103/PhysRevD.79.076004}
  {Phys. Rev. D \textbf{79} (2009), 076004}.

\bibitem{Bediaga:2003zh}
  I.~Bediaga, F.~S.~Navarra and M.~Nielsen,
  The Structure of f0(980) from charmed mesons decays,
  \href{https://doi.org/10.1016/j.physletb.2003.10.102}
  {Phys. Lett. B \textbf{579} (2004), 59-66}.

\bibitem{Aliev:2007uu}
  T.~M.~Aliev and M.~Savci,
  Semileptonic decays of pseudoscalar mesons to scalar $f_0$ meson,
  \href{https://doi.org/10.1209/0295-5075/90/61001}
  {EPL \textbf{90} (2010), 61001}.

\bibitem{Bediaga:2003hr}
  I.~Bediaga and M.~Nielsen,
  $D_s$ decays into $\phi$ and $f_0(980)$-mesons,
  \href{https://doi.org/10.1103/PhysRevD.68.036001}
  {Phys. Rev. D \textbf{68} (2003), 036001}.

\bibitem{Cheng:2023knr}
  S.~Cheng and S.~L.~Zhang,
  $D_s \rightarrow f_0(980)$ form factors and the $D_s^+ \rightarrow (f_0(980) \rightarrow )\left[ \pi \pi \right] _{S}^{\textrm{I} = 0} e^+ \nu _e$ decay from light-cone sum rules,
  \href{https://doi.org/10.1140/epjc/s10052-024-12734-5}
  {Eur. Phys. J. C \textbf{84} (2024), 379}.

\bibitem{Braun:1988qv}
   V.~M.~Braun and I.~E.~Filyanov,
   \textit{QCD Sum Rules in Exclusive Kinematics and Pion Wave Function},
   \href{https://doi.org/10.1007/BF01548594}
    {Z. Phys. C {\bf 44} (1989) 157}.

\bibitem{Balitsky:1989ry}
   I.~I.~Balitsky, V.~M.~Braun and A.~V.~Kolesnichenko,
   \textit{Radiative Decay $\Sigma^+\to p\gamma$ in Quantum Chromodynamics},
   \href{https://doi.org/10.1016/0550-3213(89)90570-1}
    {Nucl. Phys. B {\bf 312} (1989) 509}.

\bibitem{Chernyak:1990ag}
   V.~L.~Chernyak and I.~R.~Zhitnitsky,
   \textit{B meson exclusive decays into baryons},
   \href{https://doi.org/10.1016/0550-3213(90)90612-H}
    {Nucl. Phys. B {\bf 345} (1990) 137}.

\bibitem{Ball:1991bs}
   P.~Ball, V.~M.~Braun and H.~G.~Dosch,
   \textit{Form-factors of semi-leptonic D decays from QCD sum rules},
   \href{https://doi.org/10.1103/PhysRevD.44.3567}
    {Phys. Rev. D {\bf 44} (1991) 3567}.

\bibitem{Cheng:2005nb}
  H.~Y.~Cheng, C.~K.~Chua and K.~C.~Yang,
  Charmless hadronic B decays involving scalar mesons: Implications to the nature of light scalar mesons,
  \href{https://doi.org/10.1103/PhysRevD.73.014017 }
  {Phys. Rev. D \textbf{73} (2006), 014017}.

\bibitem{DeFazio:2001uc}
  F.~De Fazio and M.~R.~Pennington,
  Probing the structure of $f_0(980)$ through radiative $\phi$ decays,
  \href{https://doi.org/10.1016/S0370-2693(01)01200-X }
  {Phys. Lett. B \textbf{521} (2001), 15-21}.

\bibitem{Govaerts:1983ka}
    J.~Govaerts, F.~de Viron, D.~Gusbin and J.~Weyers,
    \textit{Exotic mesons from QCD sum rules},
    \href{https://doi.org/10.1016/0370-2693(84)92038-0}
    {Phys. Lett. B {\bf 128} (1983) 262}.

\bibitem{Govaerts:1984bk}
    J.~Govaerts, F.~de Viron, D.~Gusbin and J.~Weyers,
    \textit{QCD Sum Rules and Hybrid Mesons},
    \href{https://doi.org/10.1016/0550-3213(84)90583-2}
    {Nucl. Phys. B {\bf 248} (1984) 1}.

\bibitem{Huang:1989gv}
    T.~Huang and Z.~Huang,
    \textit{Quantum Chromodynamics in Background Fields},
    \href{https://doi.org/10.1103/PhysRevD.39.1213}
    {Phys. Rev. D {\bf 39} (1989) 1213}.

\bibitem{Shifman:1978bx}
    M.~A.~Shifman, A.~I.~Vainshtein and V.~I.~Zakharov,
    \textit{QCD and Resonance Physics. Theoretical Foundations},
    \href{https://doi.org/10.1016/0550-3213(79)90022-1}
    {Nucl. Phys. B {\bf 147} (1979) 385}.

\bibitem{Shifman:1978by}
    M.~A.~Shifman, A.~I.~Vainshtein and V.~I.~Zakharov,
    \textit{QCD and Resonance Physics: Applications}
    \href{https://doi.org/10.1016/0550-3213(79)90023-3}
    {Nucl. Phys. B\textbf{147} (1979) 448}.

\bibitem{Wang:2016wpc}
  W.~Wang,
  Search for the $a_0(980)-f_0(980)$ mixing in weak decays of $D_s/B_s$ mesons,
  \href{https://doi.org/10.1016/j.physletb.2016.06.007}
  {Phys. Lett. B \textbf{759} (2016), 501-506}.

\bibitem{Flatte:1976xv}
  S.~M.~Flatte,
  On the Nature of 0+ Mesons,
  \href{https://doi.org/10.1016/0370-2693(76)90655-9}
  {Phys. Lett. B \textbf{63} (1976), 228-230}.

\bibitem{LHCb:2014ooi}
  R.~Aaij \textit{et al.} [LHCb],
  Measurement of resonant and CP components in $\bar{B}_s^0\to J/\psi\pi^+\pi^-$ decays,
  \href{https://doi.org/10.1103/PhysRevD.89.092006}
  {Phys. Rev. D \textbf{89} (2014), 092006}.

\bibitem{Achasov:2020qfx}
  N.~N.~Achasov, A.~V.~Kiselev and G.~N.~Shestakov,
  Semileptonic decays $D\to\pi^+\pi^-e^+\nu_e$ and $D_s\to\pi^+\pi^-e^+\nu_e$ as the probe of constituent quark-antiquark pairs in the light scalar mesons,
  \href{https://doi.org/10.1103/PhysRevD.102.016022}
  {Phys. Rev. D \textbf{102} (2020), 016022}.

\bibitem{Sun:2010nv}
  Y.~J.~Sun, Z.~H.~Li and T.~Huang,
  $B_{(s)}\to S$ transitions in the light cone sum rules with the chiral current,
  \href{https://doi.org/10.1103/PhysRevD.83.025024}
  {Phys. Rev. D \textbf{83} (2011), 025024}.

\bibitem{Artuso:2008vf}
  M.~Artuso, B.~Meadows and A.~A.~Petrov,
  Charm Meson Decays,
  \href{https://doi.org/10.1146/annurev.nucl.58.110707.171131}
  {Ann. Rev. Nucl. Part. Sci. \textbf{58} (2008), 249-291}.


\bibitem{BES:2004twe}
  M.~Ablikim \textit{et al.} [BES Collaboration],
  Resonances in $J/\Psi\to\phi\pi^+\pi^-$ and $\phi K^+ K^-$,
  \href{https://doi.org/10.1016/j.physletb.2004.12.041}
  {Phys. Lett. B \textbf{607} (2005), 243-253}.

\bibitem{Duplancic:2008ix}
  G.~Duplancic, A.~Khodjamirian, T.~Mannel, B.~Melic and N.~Offen,
  Light-cone sum rules for $B\to\pi$ form factors revisited,
  \href{https://doi.org/10.1088/1126-6708/2008/04/014}
  {JHEP \textbf{04} (2008), 014}.

\bibitem{Belyaev:1993wp}
  V.~M.~Belyaev, A.~Khodjamirian and R.~Ruckl,
  QCD calculation of the $B\to\pi, K$ form-factors,
  \href{https://doi.org/10.1007/BF01474633}
  {Z. Phys. C \textbf{60} (1993), 349-356}.

\bibitem{Maiani:2004uc}
  L.~Maiani, F.~Piccinini, A.~D.~Polosa and V.~Riquer,
  A New look at scalar mesons,
  \href{https://doi.org/10.1103/PhysRevLett.93.212002}
  {Phys. Rev. Lett. \textbf{93} (2004), 212002}.

\bibitem{Kim:2018zob}
  H.~Kim, K.~S.~Kim, M.~K.~Cheoun, D.~Jido and M.~Oka,
  Further signatures to support the tetraquark mixing framework for the two light-meson nonets,
  \href{https://doi.org/10.1103/PhysRevD.99.014005}
  {Phys. Rev. D \textbf{99} (2019), 014005}.


\bibitem{Cheng:2002ai}
  H.~Y.~Cheng,
  Hadronic $D$ decays involving scalar mesons,
  \href{https://doi.org/10.1103/PhysRevD.67.034024}
  {Phys. Rev. D \textbf{67} (2003), 034024}.

\bibitem{Gokalp:2004ny}
  A.~Gokalp, Y.~Sarac and O.~Yilmaz,
  An Analysis of $f_0-\sigma$ mixing in light cone QCD sum rules,
  \href{https://doi.org/10.1016/j.physletb.2005.01.055 }
  {Phys. Lett. B \textbf{609} (2005), 291-297}.

\bibitem{Braun:2003rp}
  V.~M.~Braun, G.~P.~Korchemsky and D.~M\"uller,
  The Uses of conformal symmetry in QCD,
  \href{https://doi.org/10.1016/S0146-6410(03)90004-4 }
  {Prog. Part. Nucl. Phys. \textbf{51} (2003), 311-398}.

\bibitem{Hu:2021zmy}
  D.~D.~Hu, H.~B.~Fu, T.~Zhong, L.~Zeng, W.~Cheng and X.~G.~Wu,
  $\eta ^{(\prime )}$-meson twist-2 distribution amplitude within QCD sum rule approach and its application to the semi-leptonic decay $ D_s^+ \to \eta ^{(\prime )}\ell ^+ \nu _\ell $,
  \href{https://doi.org/10.1140/epjc/s10052-021-09958-0}
  {Eur. Phys. J. C \textbf{82} (2022), 12}.

\bibitem{Zhong:2021epq}
  T.~Zhong, Z.~H.~Zhu, H.~B.~Fu, X.~G.~Wu and T.~Huang,
  Improved light-cone harmonic oscillator model for the pionic leading-twist distribution amplitude,
  \href{https://doi.org/10.1103/PhysRevD.104.016021 }
  {Phys. Rev. D \textbf{104} (2021), 016021}.

\bibitem{Hu:2023pdl}
  D.~D.~Hu, X.~G.~Wu, H.~B.~Fu, T.~Zhong, Z.~H.~Wu and L.~Zeng,
  Properties of the $\eta _q$ leading-twist distribution amplitude and its effects to the $B/D^+ \rightarrow \eta ^{(\prime )}\ell ^+ \nu _\ell $ decays,
  \href{https://doi.org/10.1140/epjc/s10052-023-12333-w}
  {Eur. Phys. J. C \textbf{84} (2024), 15}.

\bibitem{Huang:2022xny}
  D.~Huang, T.~Zhong, H.~B.~Fu, Z.~H.~Wu, X.~G.~Wu and H.~Tong,
  $K_0^*(1430)$ twist-2 distribution amplitude and $B_s,D_s \to K_0^*(1430)$ transition form factors,
  \href{https://doi.org/10.1140/epjc/s10052-023-11851-x }
  {Eur. Phys. J. C \textbf{83} (2023), 680}.

\bibitem{Wu:2022qqx}
  Z.~H.~Wu, H.~B.~Fu, T.~Zhong, D.~Huang, D.~D.~Hu and X.~G.~Wu,
  $a_0(980)$-meson twist-2 distribution amplitude within the QCD sum rules and investigation of $D\to a_0(980)(\to\eta\pi)e^+\nu_e$,
  \href{https://doi.org/10.1016/j.nuclphysa.2023.122671 }
  {Nucl. Phys. A \textbf{1036} (2023), 122671}.

\bibitem{Yang:1993bp}
  K.~C.~Yang, W.~Y.~P.~Hwang, E.~M.~Henley and L.~S.~Kisslinger,
  QCD sum rules and neutron proton mass difference,
  \href{https://doi.org/10.1103/PhysRevD.47.3001 }
  {Phys. Rev. D \textbf{47} (1993), 3001-3012}.

\bibitem{Duplancic:2015zna}
  G.~Duplancic and B.~Melic,
  Form factors of $B, B_s\to\eta^{(')}$ and $D, D_s\to\eta^{(')}$ transitions from QCD light-cone sum rules,
  \href{https://doi.org/10.1007/JHEP11(2015)138}
  {JHEP \textbf{11} (2015), 138}.

\bibitem{BESIII:2016tqo}
  M.~Ablikim \textit{et al.} [BESIII Collaboration],
  Amplitude analysis of the $\chi_{c1} \to \eta\pi^+\pi^-$ decays,
  \href{https://doi.org/10.1103/PhysRevD.95.032002 }
  {Phys. Rev. D \textbf{95} (2017), 032002}.

\bibitem{Bharucha:2010im}
  A.~Bharucha, T.~Feldmann and M.~Wick,
  Theoretical and Phenomenological Constraints on Form Factors for Radiative and Semi-Leptonic B-Meson Decays,
  \href{https://doi.org/10.1007/JHEP09(2010)090 }
  {JHEP \textbf{09} (2010), 090}.

\bibitem{Becirevic:1999kt}
  D.~Becirevic and A.~B.~Kaidalov,
  Comment on the heavy $\to$ light form-factors,
  \href{https://doi.org/10.1016/S0370-2693(00)00290-2 }
  {Phys. Lett. B \textbf{478} (2000), 417-423}.

\end{thebibliography}
\end{document}